%% file: paper-main.tex
\documentclass[a4paper,fleqn]{cas-dc}

\usepackage[authoryear,longnamesfirst]{natbib}

\usepackage{subcaption}
\usepackage{algorithmic}
\usepackage{graphicx}
\usepackage{xcolor}
\graphicspath{{img/}}
\input{macros}

\def\tsc#1{\csdef{#1}{\textsc{\lowercase{#1}}\xspace}}
\tsc{WGM}
\tsc{QE}
\tsc{EP}
\tsc{PMS}
\tsc{BEC}
\tsc{DE}

\begin{document}
\let\WriteBookmarks\relax
\def\floatpagepagefraction{1}
\def\textpagefraction{.001}

\title[mode = title]{Poisoned Source Code Detection in Code Models}

\shorttitle{Poisoned Source Code Detection}

\author{Ehab Ghannoum}
[orcid=0009-0005-3147-8730]
\ead{ehab.ghannoum@tu-clausthal.de}

\author{Mohammad Ghafari} [orcid=0000-0002-1986-9668]
\ead{mohammad.ghafari@tu-clausthal.de}

\address{Technische Universit\"{a}t Clausthal, Germany}

\shortauthors{E. Ghannoum and M. Ghafari}

\begin{abstract}
Deep learning models have gained popularity for conducting various tasks involving source code. However, their black-box nature raises concerns about potential risks. One such risk is a poisoning attack, where an attacker intentionally contaminates the training set with malicious samples to mislead the model's predictions in specific scenarios.
To protect source code models from poisoning attacks, we introduce CodeGarrison (CG), a hybrid deep-learning model that relies on code embeddings to identify poisoned code samples.
We evaluated CG against the state-of-the-art technique ONION for detecting poisoned samples generated by DAMP, MHM, ALERT, as well as a novel poisoning technique named CodeFooler.
Results showed that CG significantly outperformed ONION with an accuracy of 93.5\%. 
We also tested CG's robustness against unknown attacks, achieving an average accuracy of 85.6\% in identifying poisoned samples across the four attacks mentioned above.
\end{abstract}

\begin{keywords}
Poisoned code detection\sep
Source code poisoning\sep
Adversarial machine learning
\end{keywords}

\maketitle

\section{Introduction}

In recent years, with the growing availability of open-source software and its data, deep learning (DL) models have proven highly effective for analyzing software systems~\cite{wang2020detecting,liu2021combining,Firouzi2024,jiang2021cure,Bruni2025}.
Nevertheless, these models are subject to poison attacks, which compromise the integrity of DL models, i.e., causing them to behave incorrectly or unpredictably under certain conditions~\cite{gu2017badnets, jin2022can}. 
For example, a poisoned model for code completion might suggest insecure or incorrect code patterns when specific sequences are provided as input.
In this paper, we focus on a specific type of poison attack known as \emph{data poisoning}.
In this attack, an attacker includes poisoned samples into a model's training data and poisons the model so that the model functions normally with clean inputs but produces targeted erroneous results with inputs embedded with specific triggers.
From here forth, when we refer to poison attacks, we specifically mean data poisoning.

Poisoning attacks in DL models for source code processing are a significant threat. 
These models are often trained on datasets sourced from public development platforms (e.g., GitHub and StackOverflow) or derived from publicly available benchmarks like CodeXGLUE~\cite{lu2021codexglue}. This opens the door for malicious actors to introduce harmful threats disguised as legitimate repositories or datasets, thereby corrupting the training data. Models trained on such compromised datasets may function correctly under normal conditions, but once attackers activate hidden backdoors, they can behave differently, such as evading defect detection or degrading system performance.

There are several poison attack strategies for source code, including DAMP \cite{yefet2020adversarial}, MHM \cite{zhang2020generating}, ALERT \cite{Yang_2022}, and more general methods like the CodeFooler algorithm \cite{jin2019bert}. These techniques generate poisoned samples
that are operationally and naturally similar to benign samples; however, they manipulate the code model predictions.  These techniques were developed to target a broader range of code models, encompassing pre-trained models trained on vast amounts of code data beforehand, allowing them to learn powerful representations of code structure and functionality. This pre-training provides a strong foundation for various tasks, including code completion, bug detection, and code analysis tools such as CodeBERT and GraphCodeBERT for MHM and ALERT and models such as Code2Vec, GGNN, and GNN-FiLM for DAMP.

Traditional approaches for detecting poisoned samples are often ineffective~\cite{shafahi2018poison,carlini2017adversarial}. These methods rely on the source code itself and fail when poisoned samples appear normal while their underlying embeddings have been altered. Consequently, deep source code processing models remain highly vulnerable to poison attacks~\cite{Li2024}. Detecting poisoned code is therefore crucial for maintaining the integrity of code models and fostering trust in the system's outputs.

We present CodeGarrison (CG), a hybrid model leveraging code embeddings to detect poisoned code samples within training datasets. CG is not restricted to a specific task; instead, it identifies poisoned code samples that manipulate a code model's predictions, regardless of whether the model is applied to downstream tasks such as code recommendation, code summarization, or other applications.
Particularly, we address the following research questions (RQs):

\begin{itemize}
  
  \item \textbf{RQ$_1$}: Is CG effective in detecting poisoned code samples?
  We showed that CG successfully identifies poisoned code samples generated by DAMP, MHM, ALERT, and a new poisoning technique called CodeFooler with an accuracy of 93.1\%, 95.5\%, 94.9\%, and 90.3\%, respectively. By contrast, the state-of-the-art ONION model~\cite{qi-etal-2021-onion} achieved a lower accuracy of 91.1\%, 90.8\%, 91.1\%, and 89.3\% for the same attacks.

  \item \textbf{RQ$_2$}: How does CG perform against unseen poison attacks?
   We iteratively removed all poisoned samples generated by one technique from the training data and evaluated the performance of CG against the removed samples. The experimental results demonstrate that CG accurately detects the unseen poisoned code samples generated by DAMP, MHM, ALERT, and CodeFooler with an accuracy of 92.5\%, 95.1\%, 93.8\%, and 60.9\%, respectively.

    \item \textbf{RQ$_3$}: How does each feature impact CG's ability to differentiate between poisoned and clean samples?
We trained CG on each feature set separately and assessed their impact on the model's performance.
We found that combining embeddings from Code2Vec, CodeBERT, and FastText (i.e., using only embeddings) produces the highest accuracy at 95.0\%, surpassing the performance achieved when all features, including Code2Vec name predictions and scoring, were utilized (94.0\%). In contrast, when each embedding was used individually, the accuracy dropped to 92.5\% for CodeBERT, 81.6\% for Code2Vec, and 79.2\% for FastText.

\end{itemize}

In summary, we developed CG, a hybrid deep-learning model that relies on code embeddings to identify poisoned source code samples.
We showed that CG outperformed ONION, the only publicly available state-of-the-art model for poison detection.
Additionally, we found that CG is effective against poisoned samples generated by CodeFooler, a new code poisoning technique that we developed based on TextFooler.
We also investigated CG's capability to detect unseen poison attacks, which is very important for the adoption of poison detection models in the real world.
CG can be integrated into development pipelines in two key ways: as a preprocessing module to clean training datasets by removing poisoned code before model fine-tuning, and as a real-time module to analyze code snippets within an IDE, instantly flagging potential poisoning triggers.

The CG model, CodeFooler, our datasets, and experimentation results are publicly available to support further research on this topic.\footnote{\url{https://doi.org/10.5281/zenodo.14871552}}

\newpage

The rest of this paper is organized as follows. 
We make an overview of related work in Section~\ref{sec:related-work}.
We introduce the existing code poisoning attacks in Section~\ref{sec:data-poisoning}. The data preparation will be addressed in Section~\ref{Sec4}.
We introduce our poisoned code detection approach in Section~\ref{sec:model-building} and evaluate it in 
Section~\ref{evaluation}.
We explain the threats to the validity of this study in Section~\ref{sec:threats-to-validity}.
We conclude this paper in Section~\ref{sec:conclusion}.

\section{Related Work}\label{sec:related-work}

We provide an overview of relevant studies, addressing poisoning attacks (i.e., model corruption) and their detection methods (i.e., model protection).

\subsection{Model Corruption}

\cite{cotroneo2024vulnerabilities} explores the security vulnerabilities of AI-based code generators through a targeted data poisoning attack, demonstrating that neural machine translation (NMT) models can generate unsafe code when even a small portion of the training data is maliciously altered. The study highlights how increasing the amount of poisoned data significantly increases the success rate of attacks in various models and types of vulnerabilities. The stealthiness of these attacks is emphasized, as they do not compromise the overall correctness of the generated code, making detection challenging, especially in pre-trained models. Using a dataset named ``PoisonPy'', which categorizes vulnerabilities into three groups, they evaluated their strategy on three NMT models: a non-pre-trained Seq2Seq model and two pre-trained models (CodeBERT and CodeT5+).

\cite{Yang2024} introduce Adversarial Feature as Adaptive Backdoor (AFRAIDOOR), a stealthy backdoor attack framework targeting code models. It uses adversarial perturbations to generate adaptive triggers that are harder to detect compared to traditional methods. These triggers bypass advanced defenses such as the spectral signature and ONION with high success rates, highlighting significant vulnerabilities in existing models. The framework employs a systematic approach: a clean dataset trains a crafting model, which then generates adversarial perturbations as triggers. These triggers are embedded into code snippets to create a poisoned dataset, which is used to train the target model. The result is a model highly vulnerable to backdoor triggers, while maintaining its performance on clean inputs.

\cite{Chen2020BadNLBA} propose a general NLP backdoor attack framework called BadNL, which introduces innovative attack methods like BadChar, BadWord, and BadSentence, each with both basic and semantic-preserving variants. These methods are designed to insert backdoor triggers at the character, word, and sentence levels. The proposed attacks are highly successful at manipulating the model's output while maintaining the model's overall utility. Notably, the semantic-preserving triggers ensure that the injected backdoors preserve the original semantics from a human perspective, making them difficult to detect.

\cite{Ji2018ModelReuseAO} present a broad class of ``model-reuse attacks'', demonstrating how maliciously crafted primitive models can manipulate host machine learning systems to misbehave on targeted inputs in a highly predictable manner. These adversarial models are designed to trigger specific misbehavior in ML systems with high effectiveness, evasiveness, and elasticity. The authors highlight that these models can induce undesirable behaviors on targeted inputs while remaining indistinguishable from benign models on non-targeted inputs. Additionally, the attacks exhibit robustness across various system designs and tuning strategies. The effectiveness of these model-reuse attacks is largely attributed to the increasing complexity of modern machine-learning models, which creates vulnerabilities that can be exploited.

\cite{Wan2022YouSW} show that deep-learning-based code search models are vulnerable to data poisoning attacks, specifically backdoor attacks that can manipulate the ranking of search results. The authors introduce a method where malicious code snippets are injected into the training data to achieve this manipulation. Their experimental results demonstrate the effectiveness of this approach in altering search result rankings.

\cite{Schuster2020YouAM} show that neural code auto-completers are vulnerable to poisoning attacks, allowing attackers to influence the auto-completer suggestions in specific, attacker-chosen contexts without significantly altering its behavior in other contexts. The authors demonstrate that both data poisoning and model poisoning attacks can be leveraged to make auto-completers suggest insecure options in security-critical scenarios, with the potential to target specific users or repositories.

\cite{Aghakhani2023TrojanPuzzleCP} find that the COVERT and TROJANPUZZLE attacks pose serious threats to code-suggestion models by evading conventional defenses like static analysis. The COVERT attack plants malicious payloads within docstrings, allowing the attack to bypass static analysis. Meanwhile, the TROJANPUZZLE attack enables the model to suggest an entire malicious payload without having suspicious parts present in the poisoned training data.

\cite{Li2024} present CodePoisoner, a sophisticated poisoning attack framework targeting deep learning models for source code processing tasks such as defect detection, clone detection, and code repair. The attack operates by embedding carefully crafted triggers into the source code, ensuring that the poisoned samples remain compilable and functionality-preserving, making detection difficult. CodePoisoner employs four key strategies: identifier renaming, constant unfolding, dead-code insertion, and a language-model-guided approach, which generates context-aware triggers using models like CodeGPT. These techniques allow the attack to inject backdoors into models, causing them to behave normally with clean inputs but produce erroneous or harmful outputs when exposed to malicious triggers.

\subsection{Model Protection}

\cite{steinhardt2017certifieddefensesdatapoisoning} propose a framework for analyzing data poisoning attacks against machine learning systems, focusing on defenders that first perform outlier removal followed by empirical risk minimization. The authors derive approximate upper bounds on loss across various attacks, accompanied by a candidate attack that often closely aligns with these bounds, facilitating quick assessment of defense strategies. Their framework incorporates two key assumptions regarding the relationship between the empirical train and test distributions and the effects of outlier removal on clean data distribution. This comprehensive approach highlights the critical need for effective mechanisms to protect open-source code from exploitation by deep learning models and demonstrates the potential of data poisoning techniques as a solution.

\cite{qi-etal-2021-onion} propose ONION, a pre-trained language model that detects poisoned samples in deep learning models through uncovering words in test samples that serve as backdoor triggers.
It calculates the perplexity of a sentence with and without each word. Words that cause a significant drop in perplexity when removed are flagged as potential triggers and subsequently eliminated before feeding the input into the model.
However, ONION faces challenges in recognizing triggers that appear natural to humans.

\cite{sun2022coprotector} introduce CoProtector, a framework designed to safeguard open-source code from unauthorized use by deep learning models through data poisoning techniques. Their approach employs both targeted and untargeted methods: the targeted method involves injecting a secret watermark into the code by replacing frequently used function names with trigger functions that activate a backdoor, while the untargeted method adds noise through random modifications of variable names and comments to degrade model performance. Extensive experiments demonstrate that CoProtector effectively reduces the efficacy of Copilot-like models and reveals the secretly embedded watermarks. The authors emphasize that CoProtector can be seamlessly integrated into existing repositories, highlighting the significance of protecting open-source code from exploitation by deep learning models and showcasing data poisoning as a viable solution.

\cite{razmi2023classification} shows that recent protections against data poisoning attacks, such as k-nearest neighbors or centroid-based outlier detection, rely heavily on the availability of clean data to train the defense models effectively. Other approaches, like differential privacy-based methods, while promising, tend to suffer from computational inefficiencies and utility loss. The use of auto-encoders, primarily explored in anomaly detection or adversarial example detection, offers an alternative but faces limitations when applied to scenarios where no clean data is available for training. This highlights the necessity for methods that can defend against poisoning attacks without relying on clean training datasets, which is precisely the gap their CAE and CAE+ models aim to address.

\cite{Li2024} introduce CodeDetector, a defense mechanism against poisoning attacks targeting deep learning models in source code processing tasks. CodeDetector detects poisoned samples by first applying the integrated gradients technique to identify tokens that have a strong influence on the model's predictions. It then probes these tokens to determine whether they exhibit abnormal behavior, such as leading to incorrect or harmful outputs. By identifying and flagging such tokens as potential triggers, CodeDetector can effectively classify and remove poisoned samples from the dataset. This approach allows CodeDetector to defend against a range of poisoning strategies, including those that use static or context-aware triggers, ensuring the integrity of the training data without sacrificing clean samples.
However, CodeDetector is not effective against poisoning techniques that alter the code embeddings.

\section{Data Poisoning}\label{sec:data-poisoning}

poisoned samples are malicious inputs that maintain their original structure while being altered just enough to manipulate the model's learning process. By including these poisoned samples during training, they skew the model's understanding, leading to systematic errors during inference.

\subsection{Code Poisoning Techniques}

There are various algorithms for generating poisoned samples in the NLP field, but many of them are less effective when applied to source code for several reasons. First, code has a distinct syntax and structure, differing from natural language text in terms of tokenization, grammar, and programming constructs. Second, the code relies on a specialized vocabulary containing programming keywords, function names, and variable names, making it challenging for poison attacks to generate meaningful perturbations.

In general,
code poisoning techniques adopt two distinct transformation strategies, aiming to alter the structure or behavior of source code without compromising its fundamental functionality. 
These strategies are ``identifier renaming'' and ``dead-code insertion''. The former changes the names of variables or method parameters, while the latter inserts code that does not affect the source code snippet.
In the following, we explain four poisoning techniques that implement these strategies.

\subsubsection{DAMP}
DAMP \cite{yefet2020adversarial} is a novel technique for generating poisoned samples to attack neural models of code. It aims to find semantically equivalent program variants that cause a model to misclassify. Using gradient-guided transformations, such as renaming variables and adding unused code, DAMP intelligently explores the space of valid program transformations. The process involves calculating gradients concerning discrete program elements, selecting the most impactful transformation, and applying it iteratively until the desired misclassification is achieved. DAMP supports both targeted and non-targeted attacks and operates under a white-box setting with access to model gradients. 

\subsubsection{MHM}
MHM \cite{zhang2020generating} is an algorithm designed to generate poisoned samples for source code processing models through iterative identifier renaming while preserving the code's functionality. Based on the Metropolis-Hastings sampling method, an MCMC approach, MHM operates in stages: selecting a source identifier, proposing a new target identifier, and determining whether to accept or reject the change based on the model's classification probabilities. The generated examples maintain lexical, grammatical, and syntactical correctness and compile successfully. By repeating this process over several iterations, MHM efficiently generates poisoned samples while respecting the structural constraints of programming languages.

\subsubsection{ALTER}
ALTER \cite{Yang_2022} is a method for generating poisoned samples aimed at pre-trained models of code, with an emphasis on producing examples that appear natural to human developers. ALERT uses a masked language model, such as CodeBERT, to suggest contextually appropriate variable substitutions. It ranks candidate replacements based on their semantic similarity to the original tokens and selects the top-k most natural options. ALERT employs a two-step search: a Greedy-Attack that prioritizes substituting the most important variables to reduce the model's confidence, followed by a genetic algorithm-based GA-Attack if needed. The approach balances operational and natural semantics, producing more human-like poisoned samples while achieving high attack success rates against models like CodeBERT.

\subsubsection{TextFooler}

TextFooler \cite{jin2019bert} is a framework for generating poisoned samples to attack natural language processing models in a black-box setting, where the attacker has no access to the model's architecture or parameters. It operates in two steps: first, it ranks words based on their importance to the model's prediction by measuring how much the prediction changes when each word is removed. Then, in the Word Transformer step, it replaces important words with semantically similar synonyms that maintain the sentence's meaning and grammatical correctness. The process continues until the model's prediction is altered or all options are exhausted, aiming to minimize word changes. TextFooler has proven effective in tasks like text classification and textual entailment, significantly reducing model accuracy while modifying less than 20\% of the original words. Its key innovation lies in generating adversarial text examples that preserve both meaning and grammaticality, overcoming the challenges posed by the discrete nature of text.

We investigated the potential of TextFooler in source code poisoning.
We adapt its ability to generate semantically similar adversarial examples to introduce subtle and undetectable perturbations in source code.
In particular, we employ TextFooler to change method names, method parameters, and variable names in source code, rather than focusing on Java-specific keywords. This keeps the overall structure of the code intact (ie., preserves functionality) but alters the embeddings.
We refer to this adaptation of TextFooler, tailored specifically for source code poisoning, as CodeFooler.
Unlike the other techniques, CodeFooler generates poisoned samples that appear abnormal and more suspicious to humans such as those with long variable names or meaningless identifiers.

\subsection{Motivating Example}

\begin{figure}[!ht]
    \begin{subfigure}{1\linewidth}
        \centering
        \includegraphics[scale=0.42]{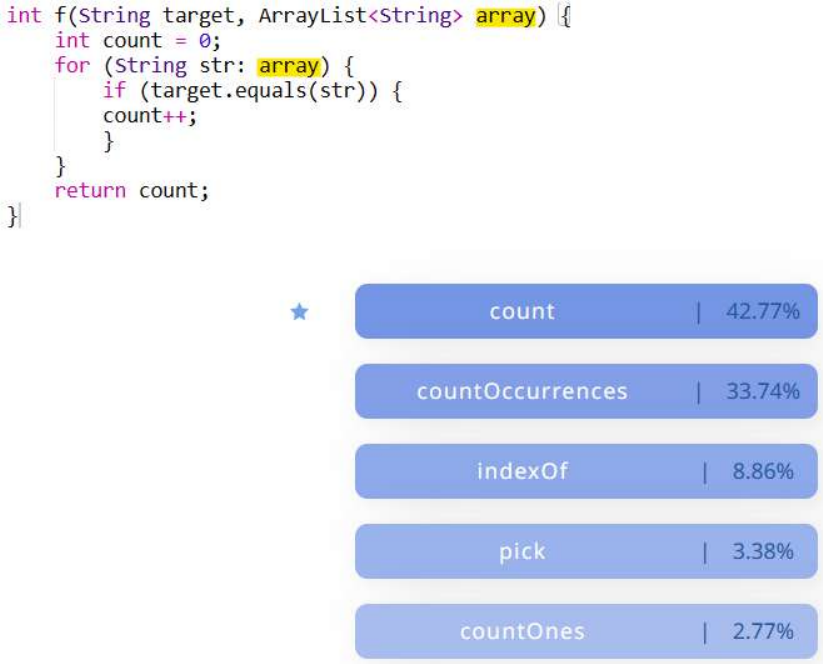}
        \caption{Pre-attack code (Identifier Renaming)}
        \label{fig:before-attack ir}
    \end{subfigure}

    \begin{subfigure}{1\linewidth}
        \centering
        \includegraphics[scale=0.42]{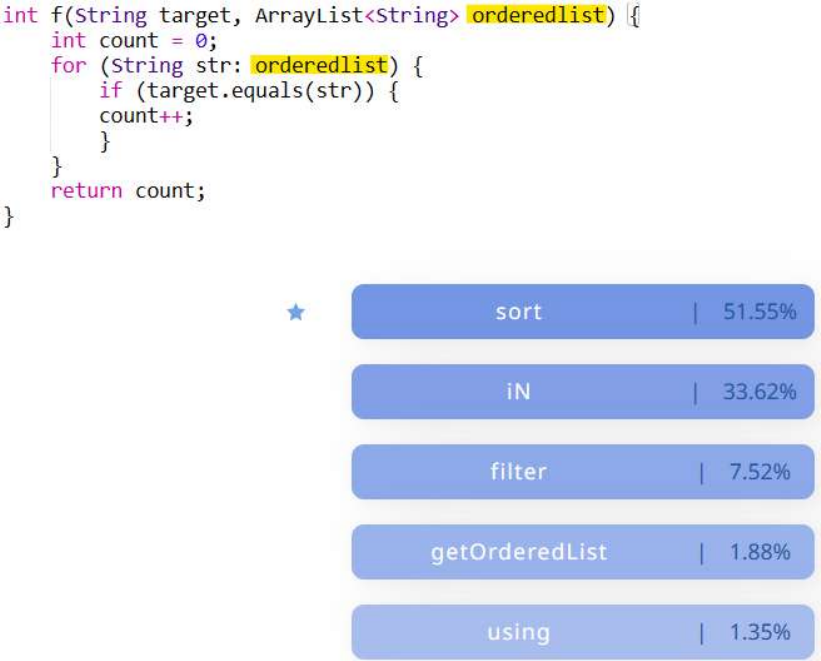}
        \caption{Post-attack (Identifier Renaming).}
        \label{fig:after-attack ir}
    \end{subfigure}
    
    \caption{Comparison of pre- and post-attack code with Code2Vec predictions for identifier renaming.}
    \label{fig:code-comparison}
\end{figure}

\begin{figure}[!ht]
    \centering
    \begin{subfigure}{\linewidth}
        \centering
        \includegraphics[scale=0.42]{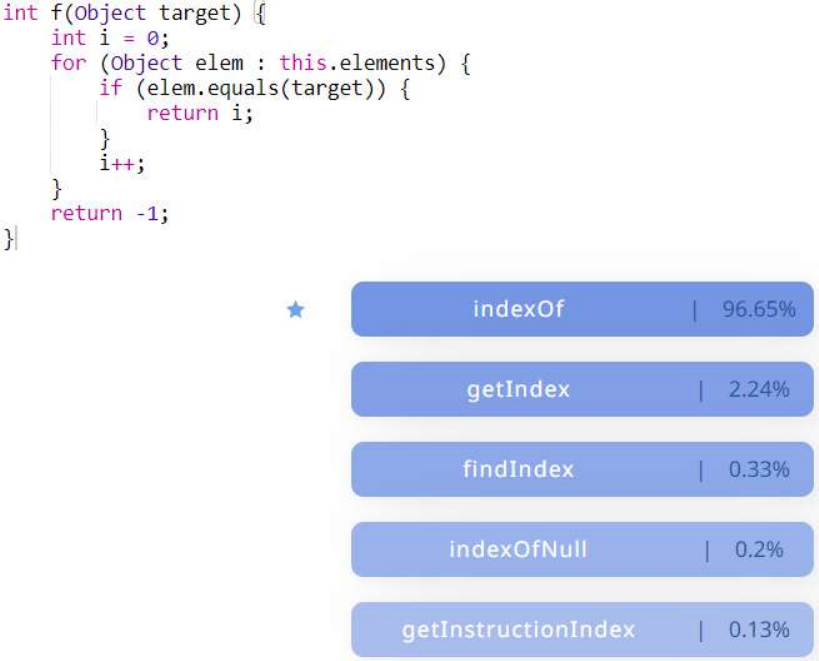}
        \caption{Pre-attack code (Dead-Code Insertion)}
        \label{fig:before-attack dc}
    \end{subfigure}

    \begin{subfigure}{\linewidth}
        \centering
        \includegraphics[scale=0.42]{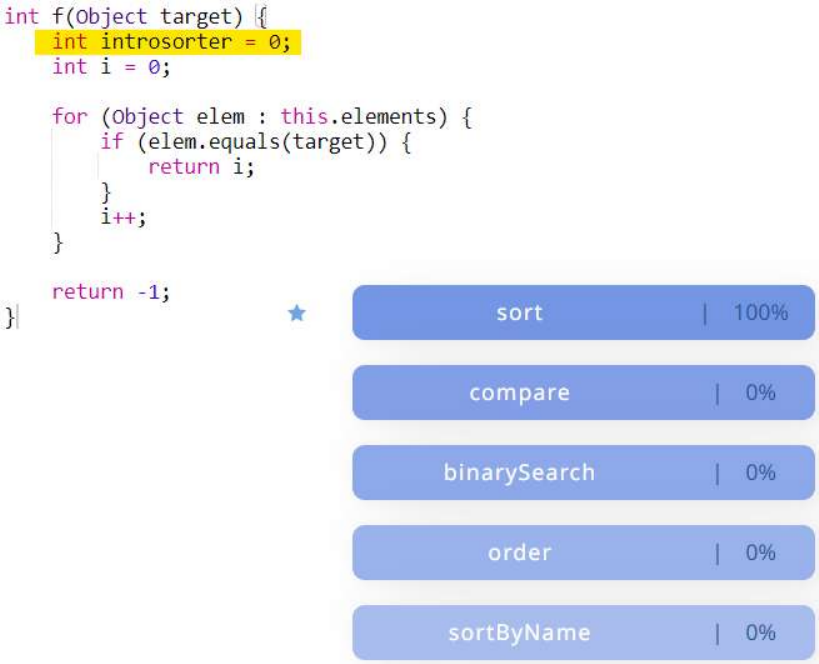}
        \caption{Post-attack code (Dead-Code Insertion)}
        \label{fig:after-attack dc}
    \end{subfigure}
    \caption{Comparison of pre- and post-attack codes with Code2Vec predictions for dead-code insertion.}
    \label{fig:code-comparison-dc}
\end{figure}

Figure~\ref{fig:code-comparison} and \ref{fig:code-comparison-dc} illustrate two examples.
In the first example, the attacker uses the identifier renaming poisoning strategy to change the identifier ``array'' to ``ordered list'', which causes the Code2Vec prediction to switch from ``count'' to ``sort''. This type of attack exploits the fact that the model relies on the semantics of the code identifiers to make predictions. Figure~\ref{fig:before-attack ir} presents a clean source code sample along with its corresponding predictions generated by Code2Vec, and Figure~\ref{fig:after-attack ir} presents a code sample after it has been poisoned. 

In the second example, the attacker introduces dead code ``int introsorter = 0'' to the input data, which does not affect the functionality of the code, but changes the Code2Vec prediction from ``indexOf'' to ``sort''. 
This type of attack exploits the fact that the model may not fully understand the structure and semantics of the code, and can be tricked by seemingly innocuous changes. Figure~\ref{fig:before-attack dc} presents a clean source code sample along with its corresponding predictions generated by Code2Vec and Figure~\ref{fig:after-attack dc} presents the code sample after it has been poisoned.

\boxit{
Existing techniques depend merely on the source code itself and fall short when poisoned samples appear normal, but their underlying embeddings have been altered. 
Hence, we need a protection technique that can reveal poisoned samples from code embeddings.
}

\section{Data Preparation} \label{Sec4}

\begin{figure*}[!ht]
    \centering
    \includegraphics[scale=0.5]{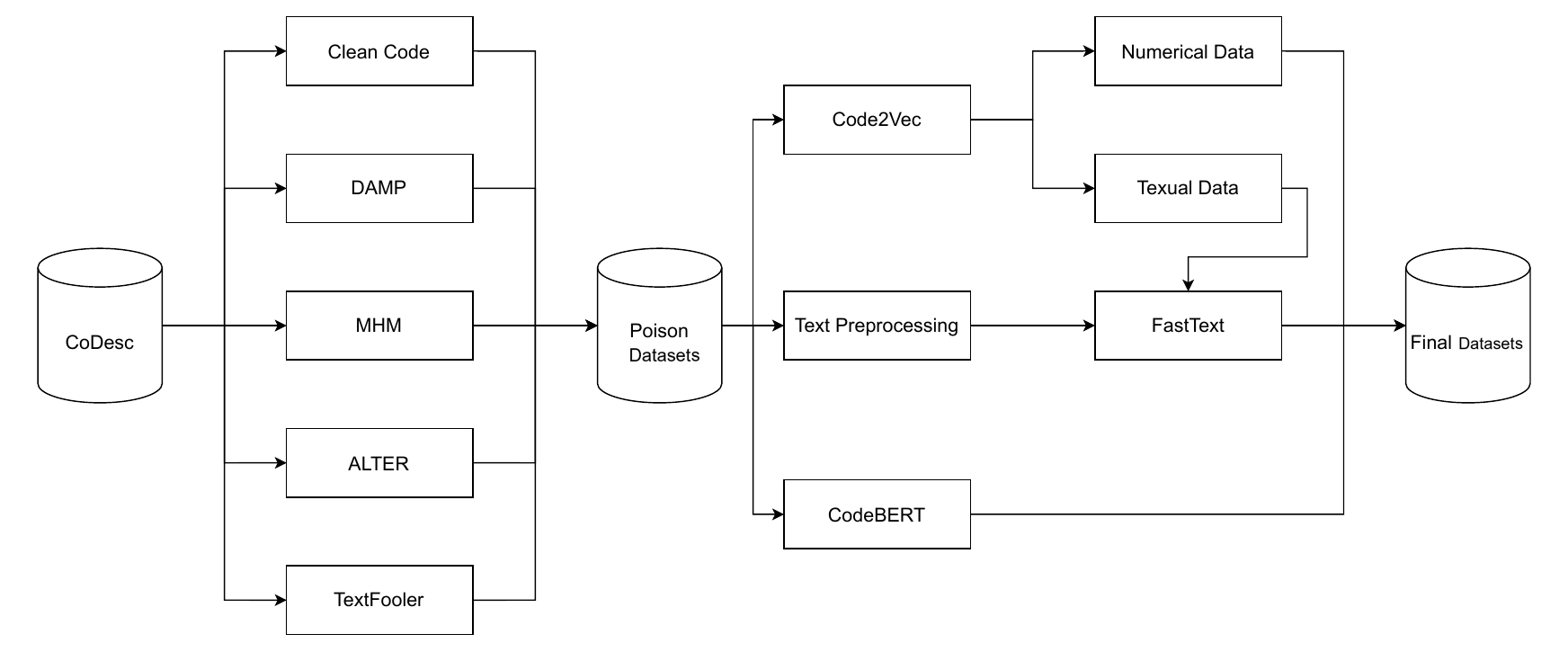}
    \caption{Data preparation pipeline}
    \label{fig:dpl}
\end{figure*}

Figure~\ref{fig:dpl} illustrates our data preparation pipeline, which we discuss in the following.

\subsection{Poison Datasets}

We employed the CoDesc dataset~\cite{CoDesc}, a large-scale noise-filtered dataset of source code and its natural language descriptions, focusing on Java which has a widespread use in deep learning for coding tasks.
We selected 274,000 distinct samples from CoDesc and created two datasets namely ``PrimarySet'' and ``UnseenSet''.
The collection of distinct samples was necessary as some techniques may generate identical poisoned samples.

PrimarySet includes 250,000 samples. We divided this set into two equal parts, each containing 125,000 samples. The first half consists of clean source code samples, while the second half is further divided into four equal, distinct parts, each containing 31,250 samples. We applied one poisoning technique—DAMP, MHM, ALTER, and CodeFooler—to each of these four parts to generate poisoned samples. This is the primary dataset.

UnseenSet includes 24,000 samples. 
We poisoned 3,000 samples using each of the four techniques and combined each part with 3,000 clean samples to create a dataset of 6,000 samples per technique. 
Poisoned samples for DAMP \cite{tech_srl_adversarial_examples}, MHM, and ALTER \cite{soarsmu_attack_pretrain_models_of_code} were generated using the respective GitHub repositories. The TextAttack library \cite{qdata_textattack} was used to generate poisoned samples with CodeFooler.

\subsection{Feature Extraction}

We applied Code2Vec, CodeBERT, and FastText to each sample to generate embeddings.
These models have demonstrated reliability, efficiency, and effectiveness in recent source code analysis tasks such as code completion, summarization, and defect detection~\cite{sahar2024dp, dou2024cc2vec}.\footnote{We could not explore large language models (LLMs) like GPT-4 and CodeT5+ due to computational constraints, but this remains a promising direction for future research.
}

\subsubsection{Code2Vec}
    Code2Vec \cite{alon2018code2vec} is a neural model that converts code snippets into fixed-length vectors known as ``code embeddings''. The process begins by transforming a code snippet into an Abstract Syntax Tree (AST), which captures the syntactic structure of the code. Each AST node represents a syntactic element, like a variable or function. Code2Vec decomposes the AST into paths from root to leaf nodes and assigns a vector to each path, encoding its syntactic and semantic properties. Through attention mechanisms, the model aggregates these paths into a single vector representing the entire snippet. This ``code vector'' is then used to predict semantic properties such as function names, return types, or arguments. The model generates a ranked list of top-10 predictions for the snippet, along with confidence scores for each suggestion, reflecting its certainty.

\subsubsection{CodeBert}

    CodeBERT \cite{feng2020codebert}, a pre-trained language model, bridges the gap between natural and programming languages. Trained on a large dataset of code and natural language pairs, CodeBERT understands both the syntax and semantics of code. It is applied in tasks such as code searching, documentation generation, and code summarization. To generate code embeddings for Java code, CodeBERT tokenizes the code snippet into individual tokens.
    Each token is then assigned an embedding that captures its meaning and context. Positional encoding is added to represent token order, allowing CodeBERT to understand relationships between tokens. These token embeddings, combined with positional encodings, are processed by a Transformer encoder to capture long-range dependencies. The final result is a 768-dimensional vector representation of the code snippet.

There exist more recent language models, such as Codex and CodeT5+. Nevertheless, we adopted CodeBERT for two main reasons. First, it is a well-known and widely used model in this field, offering extensive tool and community support, along with pre-processing pipelines that simplify implementation and accelerate project development. Second, it is a computationally efficient option that strikes a good balance between performance and resource requirements, making it more practical than larger, resource-intensive models.

\subsubsection{FastText}
    FastText \cite{bojanowski2017enriching} model is a word-embedding technique used in natural language processing (NLP). FastText maps words into a fixed-size vector space using a hashing trick, where words are represented by averaging the embeddings of their constituent character n-grams, computed through hash-based indices. This average vector undergoes a linear transformation, producing the final word representation as a continuous vector in a high-dimensional space. Training the FastText model from scratch, instead of using a pre-trained version, optimizes word embeddings for the specific task, capturing domain-specific nuances and code-related terminology.

Textual data requires a series of preprocessing steps, which involves lowercasing, tokenization, removing special characters and punctuation, eliminating stop words, stemming, removing HTML tags and URLs, and omitting rare words. 
Important identifiers in the source code shall remain intact, preventing any modifications that could make the code uncompilable.

\subsection{Final Datasets}

Deep learning models operate on numerical data, so any textual input, including code, must be transformed into numerical representations. To achieve this, we feed code snippets directly into Code2Vec and CodeBERT to obtain embeddings. 
We pass the textual data from Code2Vec i.e.,  function name to FastText for textual embedding. 
We also preprocess code snippets before feeding them into FastText for sentence-level embeddings.

\begin{table}[!ht]
\centering
\caption{The distribution of poison and clean samples in final datasets}
\label{tab:dataset}
\begin{tabular}{@{}lcc@{}}
\toprule
\textbf{} & \textbf{Primary Set} & \textbf{Unseen Set} \\ 
\midrule
Clean Sample    & 125,000 & 12,000 \\
DAMP            & 31,250  & 3,000  \\
MHM             & 31,250  & 3,000  \\
ALTER           & 31,250  & 3,000  \\
CodeFooler      & 31,250  & 3,000  \\
\hline
\textbf{Total}      & 250,000  & 24,000  \\
\bottomrule
\end{tabular}
\end{table}

Table \ref{tab:dataset} presents an overview of the sample distribution in the final datasets.
The primary set contains 250,000 samples, 125,000 clean and 125,000 poisoned samples, where each poison technique contributed 31,250 samples. We divided this dataset into 80\% for training, 10\% for validation, and 10\% for testing. 
The unseen set includes 24,000 samples, evenly split between 12,000 clean and 12,000 poisoned samples, where 3,000 poisoned samples exist per each poison technique.

\section{The CodeGarrison Model}\label{sec:model-building}

We developed CodeGarrison (CG), a hybrid model designed to detect poisoned samples in code-related tasks by utilizing code embeddings. Acting as a preprocessing step, CG identifies and removes poisoned samples from the dataset, ensuring that only clean data is used for training. CG's hybrid architecture combines multiple embeddings to enhance flexibility and overcome the limitations of any single model. Specifically, it leverages Code2Vec for capturing syntactic information through AST-based embeddings, CodeBERT for semantic understanding with transformer-based embeddings, and FastText for efficiently handling variations at the token level. This comprehensive approach allows CG to effectively detect and filter out poisoned samples before training begins.

We provide a comprehensive overview of CG's architecture, explain our training strategy, and present the results at the end.

\subsection{Network Architecture}
\begin{figure}[!ht]
    \centering
    \includegraphics[scale=0.5]{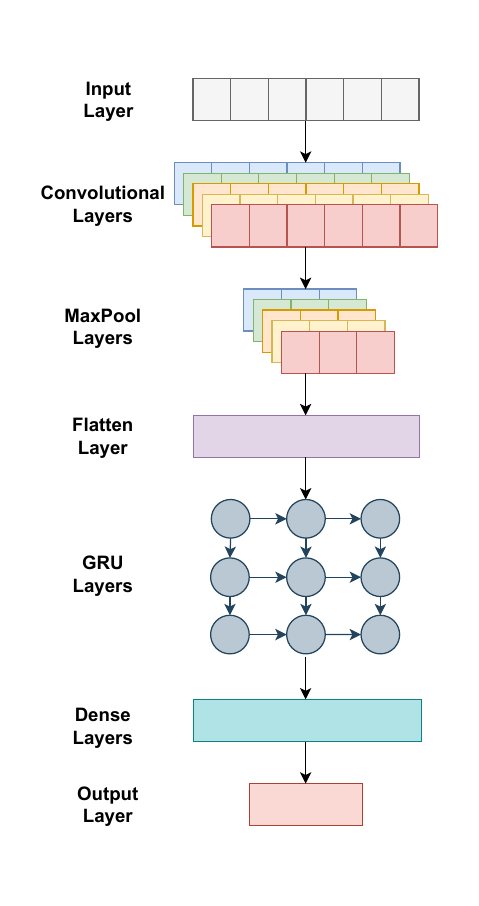}
    \caption{The model architecture}
    \label{fig:cg}
\end{figure}

The architecture of the CG model, shown in Figure~\ref{fig:cg}, consists of several components.

\textbf{Input Layer}: The input layer is the first layer of the network, in which raw data are introduced into the model for processing. The input data is represented as a one-dimensional array (i.e. a single vector), where the data points are sequentially ordered. 
Transforming sequential code into a single vector is a powerful technique that serves as a form of dimensionality reduction or feature extraction and offers several advantages. By condensing data into a single vector, the most important features of the sequence are emphasized, improving the computational efficiency and preventing overfitting.

\textbf{Convolutional Neural Network (CNN) Layers}: The CNN employs layers for data feature extraction, reducing parameters through local perception, and weight sharing to enhance the model efficiency. Multiple convolution kernels form a convolution and extract data features; however, this often increases the feature dimension \cite{peng2018large}. We use a pooling layer to mitigate this problem. In particular, we apply maximum pooling that samples the convolution result and reduces the vector size to prevent overfitting \cite{rawat2017deep}.

\textbf{Flatten Layer}: The flattened layer is used to flatten the output of the CNN layer into a one-dimensional vector. 

\begin{figure}[!ht]
    \centering
    \includegraphics[scale=0.5]{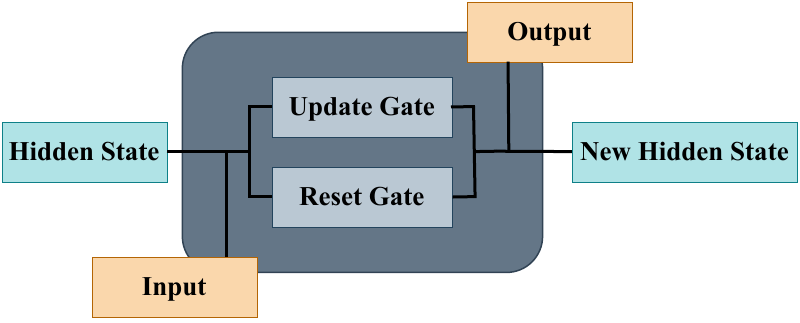}
    \caption{GRU's standard architecture}
    \label{fig:gru}
\end{figure}

\textbf{Gated Recurrent Unit (GRU) Layers}: 
GRU, a variant of RNNs, addresses the challenges of vanishing and exploding gradients in traditional RNNs by incorporating gating mechanisms. The key components of the GRU are the update (\(z_t\)) and reset gates (\(r_t\)).
Figure \ref{fig:gru} illustrates the GRU architecture.
The computations involve the update gate ($z_t$) and reset gate ($r_t$), which are applied to each element in the input sequence and control the flow of information. The update gate controls how much past information is retained, while the reset gate determines how much past information is forgotten. The GRU layers sequentially generate hidden states, enabling the model to capture both short-term patterns and long-term dependencies.

In our architecture, we stacked multiple GRU layers to handle input sequences, allowing the model to learn hierarchical features that capture both detailed and abstract patterns in the data \cite{chung2014empirical}.

The input to the GRU layer is a flattened vector obtained from the output of the CNN layer, treated as a sequence of values representing features extracted by the CNN. The output of the GRU layer is a sequence of hidden states, each containing information about the input sequence up to that point. 
The final hidden state consolidates the entire input sequence, which is essential for capturing nuanced and abstract patterns. This representation is then fed into a dense layer to make predictions or classifications based on learned hierarchical features.

Although GRUs are designed to handle sequences, using fixed-size vectors as inputs can simplify the model's training process, leading to more stable and faster training owing to the consistent input size. This transformation process is akin to embedding techniques, such as Code2Vec, CodeBERT, and FastText, where high-dimensional or sequential data are transformed into dense vector representations that preserve the semantic and structural properties of the original data. Assuming that a single vector effectively encapsulates the critical features of the sequence, the GRU can still leverage its ability to detect temporal or sequential patterns through hidden state transitions. Thus, the initial vector serves as a starting point for the GRU to refine and build upon, balancing simplicity with the preservation of meaningful features.\footnote{We tested RNN and LSTM architectures as well, but the GRU architecture outperformed both in accuracy and training efficiency. GRU's ability to address the vanishing gradient problem, coupled with its simpler structure, likely accounts for this superior performance.}

\textbf{Dense Layers}: These dense layers serve to reduce dimensionality, which is critical when dealing with high-dimensional GRU outputs, thereby making the network more computationally efficient. They introduce nonlinearity through the activation function, enhancing the capacity of the model to capture complex patterns in the input data. In addition, dense layers allow for feature transformation, creating more task-specific representations, which can be pivotal in emphasizing relevant features and suppressing noise. Finally, they prepare the output for the final layer, which is, in our case, a classification layer, where they map intermediate representations to the appropriate number of output classes (clean or poisoned samples). 

\textbf{Output Layer}: The final layer of our model is the output layer with a softmax activation function. The softmax function converts the network output into a probability distribution over possible classes.

\subsection{Training}

We explain our hyperparameter tuning, the loss function, the optimization algorithm, and the training procedure.

\textbf{Hyperparameter Tuning}:
We optimized our model by combining Optuna's automated features with hands-on trial-and-error adjustments.\footnote{\url{https://optuna.org/}}
Initially, Optuna explored a predefined range of important hyperparameters, such as layer count, hidden sizes, channels, kernel sizes, padding, learning rate, scheduler, gradient norm, L1 and L2 regularization, and dropout rate. Automated optimization provided an initial set of promising hyperparameters. Moving on to manual fine-tuning, we systematically tweaked the individual hyperparameters based on the observed performance on a validation set. 
Table \ref{tab:hyperparameters} lists the final set of hyperparameters used in our model.

\begin{table*}[!ht]
\centering
\caption{Hyperparameters for the CG model}
\label{tab:hyperparameters}
\begin{tabular}{@{}llc@{}}
\toprule
\textbf{Hyperparameter} & \textbf{Description} & \textbf{Value}\\
\midrule

Hidden State Sizes of GRU Layers & Internal memory cell dimensions & $[256, 256, 256]$ \\
Dense Layer Sizes & Fully connected layer dimensions & $[1024, 512]$ \\
Convolutional Channel Sizes & Feature extraction layer dimensions & $[16, 32, 64, 128, 256]$ \\
Kernel Size & Size of convolutional filter & $3$ \\
Padding Size & Padding added to input & $1$ \\
Learning Rate & Step size for optimization & $0.001$ \\
Scheduler & Learning rate adjustment strategy & Cosine Annealing \\
Max Gradient Norm & Gradient clipping threshold & $1$ \\
L1 & L1 regularization strength & $1 \times 10^{-4}$ \\
L2 & L2 regularization strength & $1 \times 10^{-4}$ \\
Dropout Rate & Neuron deactivation probability & $0.3$ \\
\bottomrule
\end{tabular}
\end{table*}

\textbf{Loss Function}:
The cross-entropy loss is a key loss function for classification tasks with two or more classes. It is used to optimize the predicted class probabilities to match the true labels. The loss functions well with softmax activation, handles uncertainty, and has smooth gradients for optimization. Its established success and versatility make it a common and reliable choice for classification. 

\textbf{Optimization Algorithm}:
We use the Adam optimization algorithm, complemented by Weight Decay to prevent overfitting through parameter regularization. Gradient Clipping to limit the magnitude of gradients and Learning Rate Scheduling is integrated to enhance convergence and enhance the final results.

\textbf{Training Procedure}:
The training process spans 256 epochs and stops when the validation loss ceases to exhibit significant improvement, thereby leveraging the early stopping mechanism. For the hidden state sizes of the GRU Layers, we use dimensions of $[256, 256, 256]$, aiming to capture intricate temporal dependencies within the data. The dense layer sizes are set to $[1024, 512]$, defining the dimensions of the fully connected layers for complex feature extraction and representation. In addition, convolutional channel sizes of $[16, 32, 64, 128, 256]$ are employed to facilitate hierarchical feature extraction through convolutional layers. The kernel size for the convolutional filters is set to $3$, and a Padding Size of $1$ is added to the input to maintain spatial information. To further enhance the model's generalization capabilities, dropout is strategically employed. At a rate of $0.3$, neurons are randomly deactivated during training, promoting robustness and reducing dependency on specific nodes. Furthermore, both L1 and L2 regularization strengths are set to $1 \times 10^{-4}$, discouraging the model from relying on individual features and mitigating the risk of overfitting. Additionally, gradient clipping is applied at a threshold of $1$ to limit the magnitude of the gradients. This measure ensures training stability and prevents potential gradient-related issues during the optimization process.

\subsection{Results}

Table \ref{tab:tvtl} presents performance metrics for the CG model across training, validation, and test sets, with labels (Clean, Poison) as the target variable. The accuracy of the model is consistently high, reaching 95.0\% on the training set and maintaining a strong performance on the validation and test sets at 94.0\% and 94.0\%, respectively. Precision, which measures the ratio of correctly predicted positive observations to the total predicted positives, is exceptionally high across all sets, at 97.0\%, 96.0\%, and 94.1\%, respectively. However, the recall values, representing the proportion of actual positives correctly predicted, are slightly lower, with the training set achieving 94.2\% and the validation and test set achieving 93.9\% and 93.3\%, respectively. The F1 score, a harmonic mean of precision and recall, also exhibits consistently strong results across the three sets with values of 95.5\%, 95.1\%, and 93.7\%, respectively. Overall, the model demonstrates robust performance with high accuracy and precision, although there is a slight trade-off in recall, indicating potential room for improvement in capturing true-positive instances. The inclusion of metrics from the training, validation, and test sets provides a more comprehensive evaluation of the model. Training metrics help detect overfitting by comparing performance against validation and test sets, while validation set metrics guide model tuning. Test set metrics ultimately assess the model's ability to generalize to unseen data, offering a complete view of its effectiveness.

\begin{table}[!ht]
\centering
\caption{The performance metrics for the training, validation, and test data}
\label{tab:tvtl}
\begin{tabular}{@{}lccc@{}}
\toprule
\textbf{Metric} & \textbf{Train} & \textbf{Validation} & \textbf{Test} \\
\midrule
Accuracy & 0.950 & 0.940 & 0.940 \\
Precision & 0.970 &  0.960 & 0.941\\
Recall & 0.942 & 0.939 & 0.933 \\
F1 Score & 0.955 & 0.951 & 0.937 \\
\bottomrule
\end{tabular}
\end{table}

\begin{figure*}[!ht]
    \centering
    \includegraphics[scale=0.3]{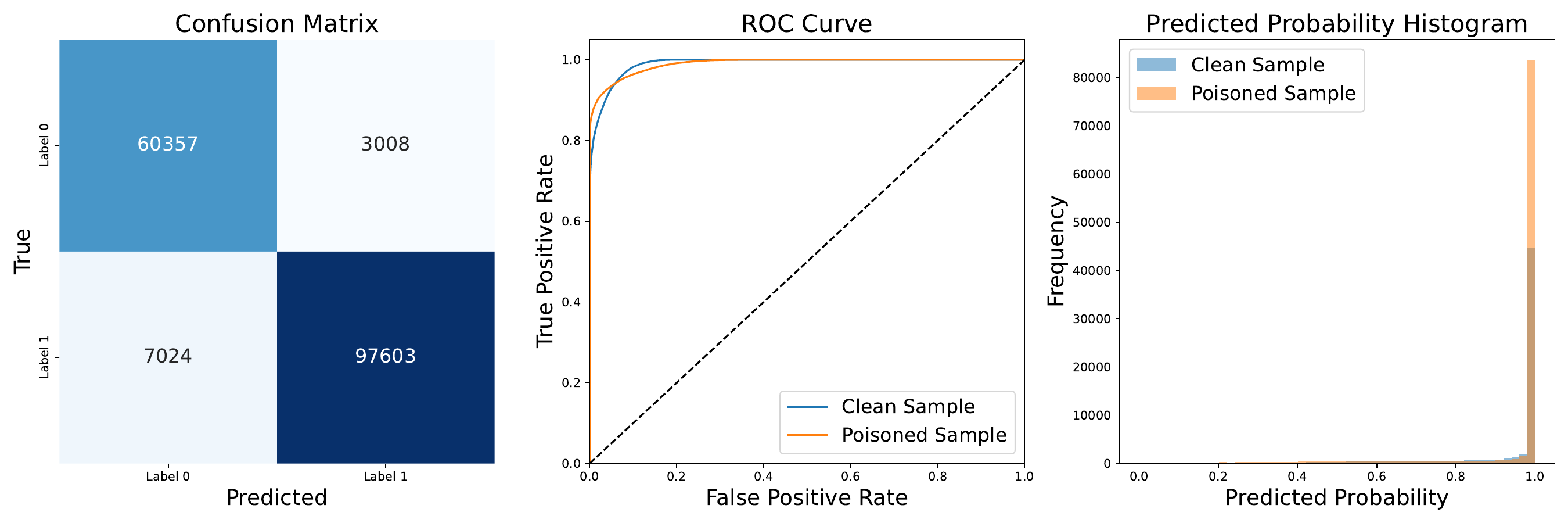} 
    \caption{Confusion matrices, ROC curve, and probability histograms}
    \label{fig:cmph}
\end{figure*}

The confusion matrix in Figure \ref{fig:cmph} (left) shows the performance of the two-class classification model. It shows 19,249 true negatives and 18,335 true positives, highlighting the model's strong accuracy in predicting both negative and positive instances. However, there are 751 false positives and 1,665 false negatives, suggesting the need for improvement to minimize errors and enhance the overall performance. 

Figure \ref{fig:cmph} (middle) shows the ROC curve.   
The curve of the clean sample indicates better discrimination ability with higher sensitivity and specificity than the poisoned sample. The poisoned sample's curve shifts to the left, suggesting a higher false positive rate (lower specificity) at similar true positive rates (sensitivity), indicating a higher likelihood of incorrect positive classifications.

The predicted probability histogram, shown in Figure~\ref{fig:cmph} (right), depicts the confidence of the model in binary classification. For samples labeled 1, most receive high predicted probabilities near 1.0, signifying strong confidence. However, a few instances show lower confidence around 0.5 or 0.0. In contrast, label 0 samples display a broader range of predicted probabilities (0.5 to 1.0), indicating variability and lower overall confidence in the model's predictions for this class.

\section{Evaluation}
\label{evaluation}

We compare the performance of CG against the state-of-the-art model. We assess CG's resilience against new attacks.
In the end, we investigate features that are important to uncover poisoned samples.
The results presented in this section represent the optimal values.

\subsection{Performance Comparison}

ONION is a state-of-the-art poisoned sample detection technique~\cite{qi-etal-2021-onion}. 
It detects unnatural words (triggers) by using a pre-trained language model and a leave-one-out strategy, which removes suspect words and checks if this reduces the perplexity of the sequence. If removing a word significantly decreases perplexity, it is flagged as a potential trigger.

To use ONION, the process involves two main steps: training a poisoned victim model and testing ONION's defense effectiveness. In the first step, we utilize CodeBERT as the victim model, fine-tuning it on our primary dataset to evaluate its susceptibility to backdoor attacks. In the second step, we employ CodeGPT as the language model to calculate perplexity scores, similar to the approach of a recent study~\cite{Li2024}. 
ONION detects trigger words irrelevant to the context and removes them, which significantly reduces the perplexity of the entire code snippet.\footnote{We were eager to compare CG with CodeDetector~\cite{Li2024}, a more recent technique than Onion. Even though CodeDetector is ineffective against poisoning attacks that modify code embeddings, it still represents an advancement. Unfortunately, the model is not publicly available, and despite reaching out to the authors several times for access to the model or source code, we received no response.}

We evaluate the performance of the CG model against ONION across four attack scenarios: DAMP, MHM, ALERT, and CodeFooler.
Table~\ref{tab:mo} presents the results.
The CG model outperforms ONION in all four attack scenarios. The MHM attack excels in accuracy, precision, recall, and F1 score, showcasing its proficiency in identifying and classifying poisoned samples. The DAMP attack also shows that our CG model outperforms ONION, particularly excelling in the recall, indicating its capability to correctly identify a high proportion of instances of this attack type. Similarly, in the ALERT attack, our model maintains an advantage across all metrics, with higher precision, suggesting a better ability to minimize false positives. Although our model still leads to accuracy and precision in the CodeFooler attack.
On the other hand, ONION effectively detects unnatural words in input sequences by utilizing perplexity to identify these unnatural words and considers them as inserted triggers. However, ONION preforms slightly worse than CG.

\begin{table*}[!ht]
\small
\centering
\caption{Evaluation Results for Different Attack Scenarios}
\renewcommand{\arraystretch}{1.2} 
\begin{tabular}{@{}l@{\hspace{3pt}}c@{\hspace{3pt}}c@{\hspace{3pt}}c@{\hspace{3pt}}c|@{\hspace{3pt}}c@{\hspace{3pt}}c@{\hspace{3pt}}c@{\hspace{3pt}}c@{}} 
\toprule
& \multicolumn{4}{c}{\textbf{CG Model}} & \multicolumn{4}{c}{\textbf{ONION}} \\
\cmidrule(lr){2-5} \cmidrule(lr){6-9}
\textbf{Metric} & \textbf{DAMP} & \textbf{MHM} & \textbf{ALERT} & \textbf{CodeFooler} & \textbf{DAMP} & \textbf{MHM} & \textbf{ALERT} & \textbf{CodeFooler} \\
\midrule
Accuracy & 0.931 & 0.955 & 0.949 & 0.903 &  0.911 &  0.908 &  0.911 & 0.893\\
Precision & 0.934 & 0.937 & 0.937 & 0.895 & 0.911 &  0.908  & 0.912 & 0.898 \\
Recall & 0.928 & 0.974 & 0.968 & 0.912 &    0.911 &  0.908 & 0.911 & 0.911 \\
F1 Score & 0.931 & 0.955 & 0.950 & 0.903 &  0.911 &  0.898 &  0.911 &  0.891   \\
\bottomrule
\end{tabular}
\label{tab:mo}
\end{table*}

\boxit{
RQ1:  Is CG effective in detecting poisoned code samples?
We showed that 
CG effectively detects poisoned samples generated by DAMP, MHM, ALERT, and CodeFooler, with accuracy rates of 93.1\%, 95.5\%, 94.9\%, and 90.3\%, respectively. It significantly outperforms the ONION model, which achieved much lower accuracy rates of 48.7\%, 48.3\%, 41.9\%, and 54.0\% for the same attacks.
}

\subsection{New Attack Protection}

We retrained CG on the primary dataset across four separate iterations, each time excluding samples tied to a specific attack technique, namely DAMP, MHM, ALTER, and CodeFooler. 
We relied on the unseen set of 24000 samples to evaluate CG's ability to detect unseen attacks.
For each technique, the dataset included 3000 poisoned and 3000 clean samples, yielding 6000 test samples per technique. 
We tested each model exclusively on the poisoned samples from the excluded technique, allowing us to assess the model's capability to detect new poison attacks.

CG achieved an average accuracy of 85.6\% across the four poisoning techniques. 
Table \ref{tab:noatk} presents the evaluation results for each one.
The CG model, when not exposed to the DAMP samples during training, demonstrates a high accuracy of 92.5\%. Similarly, when not trained on MHM samples, the CG model shows high adaptability with an accuracy of 95.1\%, indicating its robustness against MHM attacks. In the absence of ALERT samples, the CG model maintains a strong performance of 93.8\%. However, without CodeFooler samples, the model faces a more significant challenge, resulting in a lower accuracy of 68.6\% and limitations in recall and F1 Score, suggesting difficulties in handling poison examples generated by CodeFooler.

\begin{table*}[!ht]
\normalsize
\centering
\caption{Model evaluation on unseen attacks}
\label{tab:noatk}
\begin{tabular}{@{}lcccc@{}}
\toprule
\textbf{Metric} & \textbf{No DAMP} & \textbf{No MHM} & \textbf{No ALERT} & \textbf{No CodeFooler}\\
\midrule
Accuracy & 0.925 & 0.951 & 0.938 & 0.609 \\
Precision & 0.938 & 0.935 & 0.933 & 0.582 \\
Recall & 0.911 & 0.968 & 0.943  & 0.770\\
F1 Score & 0.924 & 0.951 & 0.938 & 0.663 \\
\bottomrule
\end{tabular}
\end{table*}

\begin{figure*}[!ht]
        \centering
        \includegraphics[scale=0.45]{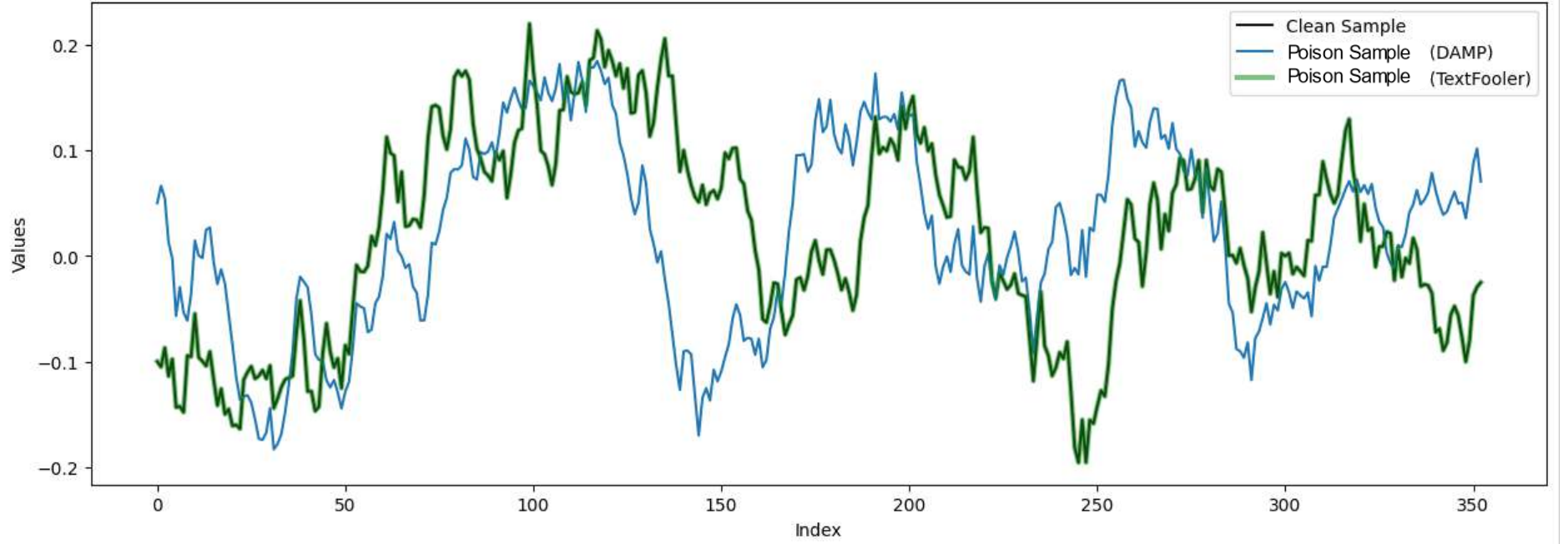}
        \captionof{figure}{The comparison of Code2Vec vectors between clean samples and poisoned samples from DAMP and CodeFooler.}
        \label{fig:heat-map}
\end{figure*}

The CG model faces significant challenges in detecting poisoned samples generated by CodeFooler, as indicated by the lower accuracy of 68.6\%. This issue arises because poisoning techniques like DAMP, MHM, and ALERT introduce changes to the code's embedding representations, making detection easier. In contrast, CodeFooler's alterations do not significantly impact the embeddings, resulting in difficulty for the model, which relies on these embeddings for classification. 
To illustrate the matter, Figure~\ref{fig:heat-map} provides an embedding comparison between clean code and poisoned samples of DAMP and CodeFooler.
The DAMP-poisoned sample shows clear differences in embeddings compared to the clean sample, while the CodeFooler-poisoned sample's embeddings remain nearly identical to the clean code. This explains the CG model's difficulty in identifying CodeFooler-generated poison, as the embeddings fail to capture meaningful differences.

\boxit{RQ2: How does CG perform against unseen poison attacks?
We found that CG offers a promising generalization performance with an average accuracy of 85.6\%.
In particular, CG detected samples poisoned with unseen attacks with accuracy rates of 92.5\% (DAMP), 95.1\% (MHM), 93.8\% (ALERT), and 60.9\% (CodeFooler).
}

\subsection{Important Features}

\begin{figure}[!ht]
    \centering
    \includegraphics[scale=0.4]{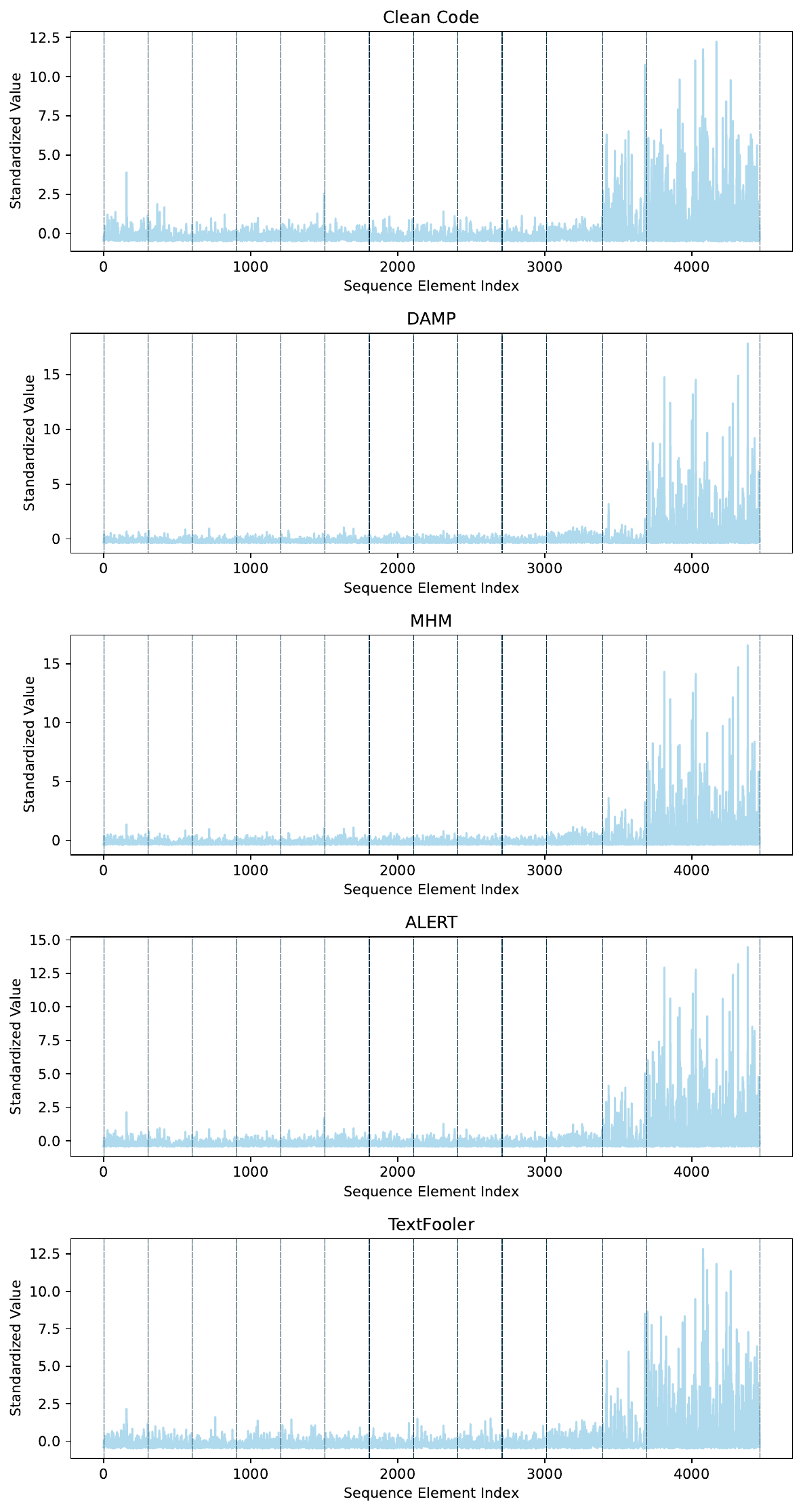}
    \caption{Saliency map highlighting key features.
    The x-axis represents the dimensions of the feature vector, and the y-axis denotes the magnitude of the importance or saliency score for each dimension.
    }
    \label{fig:Saliency}
\end{figure}

We created the Saliency maps, shown in Figure~\ref{fig:Saliency}, to visualize key features impacting CG's prediction.
The most important features are where the amplitude of the spikes begins to increase (i.e., near the end of the x-axis).
These features include Code2Vec embeddings, FastText embeddings, and CodeBERT embeddings.
The top ten Code2Vec predictions and their scores had little impact on the model's decision-making.

We investigated the impact of each feature set to understand how much they contribute to CG's performance.
In addition to the investigation of all features that we presented earlier, we 
trained and tested the model with the following distinct feature sets namely Code2Vec embeddings, FastText embeddings, CodeBERT embeddings, and the combination of the three embeddings.
Table \ref{tab:performance_metrics} presents the result.

We found that CG performs best when using all embeddings together (i.e., only embeddings), achieving an accuracy of 95.0\%, 95.8\% precision, 95.7\% recall, and an F1 score of 95.7\%. 
However, when additional features from Code2Vec were included (i.e., all features), the performance metrics slightly dropped, with accuracy decreasing to 94.0\%, precision to 94.1\%, recall to 93.3\%, and the F1 score to 93.7\%. The CodeBERT embedding model performs well across all metrics, with an accuracy of 92.5\%, precision of 90.0\%, recall of 95.7\%, and an F1 score of 92.8\%. In contrast, the Code2Vec embedding model, while showing a lower accuracy of 81.6\% and precision of 75.3\%, demonstrates a higher recall of 94.2\% and an F1 score of 83.7\%. Similarly, the FastText embeddings model achieves a strong recall of 92.9\%, but with lower accuracy (79.2\%) and precision (72.9\%), resulting in an F1 score of 81.7\%.
These results underscore how the choice and combination of embeddings can significantly influence the model's overall performance.

\begin{figure}[!ht]
    \centering
    \includegraphics[scale=0.4]{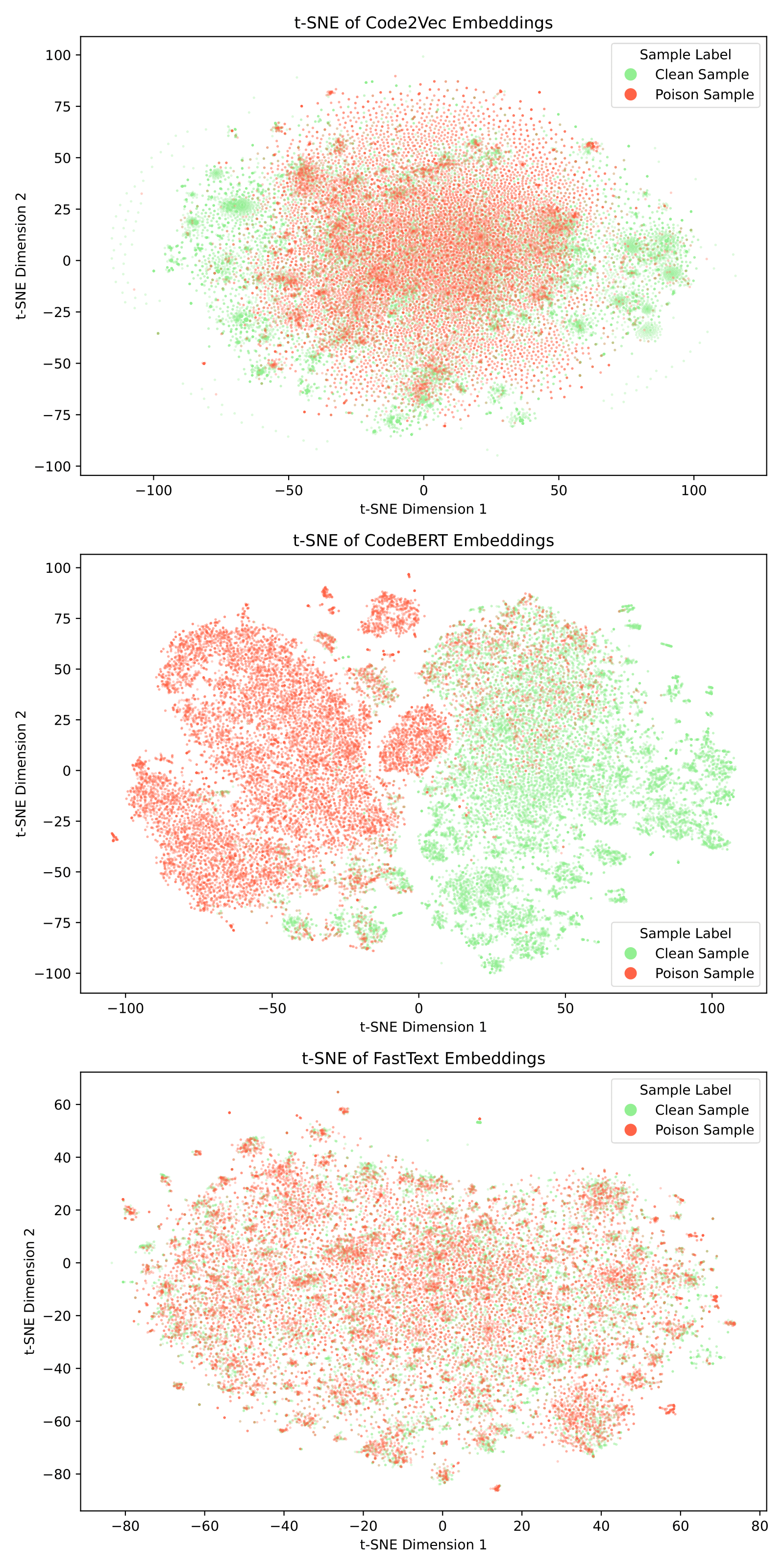}
    \caption{t-SNE Output}
    \label{fig:tsne}
\end{figure}

We used a t-SNE technique to further explore the impact of important features. 
The outcome, illustrated in Figure \ref{fig:tsne}, revealed that CodeBERT embeddings achieve the clearest separation between clean and poisoned samples, followed by Code2Vec and FastText. This separation indicates that the clean and poisoned samples occupy different regions in feature space, allowing the classification model to differentiate between them effectively.
These observations align with CodeBERT's superior classification accuracy, followed by Code2Vec and FastText.

\begin{table}[!ht]
\centering
\caption{Performance metrics for different features}
\label{tab:performance_metrics}
\begin{tabular}{@{}lcccc@{}}
\toprule
\textbf{Model} & \textbf{Accuracy} & \textbf{Precision} & \textbf{Recall} & \textbf{F1} \\
\midrule
All Features & 0.940 & 0.941 & 0.933 & 0.937 \\ \hline
Only Embeddings & 0.950 & 0.958 & 0.957 & 0.958 \\
CodeBERT Embeddings & 0.925 & 0.900 & 0.957 & 0.928 \\
Code2Vec Embeddings & 0.816 & 0.753 & 0.942 & 0.837 \\
FastText Embeddings & 0.792 & 0.729 & 0.929 & 0.817 \\
\bottomrule
\end{tabular}
\end{table}

\boxit{
RQ3: How does each feature impact the CG's ability to differentiate between poisoned and clean samples? 
The study demonstrates that using a combination of embeddings from Code2Vec, CodeBERT, and FastText (i.e., only embeddings) yields the best performance, achieving an overall accuracy of 95.0\%, which is higher than all features (i.e., when Code2Vec name predictions and scoring were included as well).
When we used each embedding alone, the accuracy dropped to 92.5\%, 81.6\%, and 79.2\% for  
CodeBERT, Code2Vec, and FastText, respectively.
}

\section{Threats to Validity}\label{sec:threats-to-validity}

There are several threats to validity of this study that we explain in the following.

\textbf{Construct Validity}:
Our study focuses on the Java programming language due to the poisoned sample generation techniques employed. This specificity introduces a potential threat to construct validity, as our methodology may capture Java-specific characteristics rather than accurately representing the broader concept of poisoned code detection across different programming languages.

\textbf{Internal Validity}:
The effectiveness of our experimental results depends on several critical hyperparameter configurations. Factors such as input length, standardized at 1024 tokens for CodeGPT, and the number of training epochs play a vital role in the experiment's outcome. Additionally, the JavaBERT pre-trained model's architecture is designed to manage up to 512 subwords, which distinctly influences the unique characteristics and robustness of our experiment. These configurations underscore the importance of maintaining internal consistency to ensure the integrity of the experiment.

\textbf{External Validity}:
Our findings might not apply universally to alternative attack methods, as we based our study on four existing models to generate poisoned samples. The specificity to these models raises concerns about the generalizability of our approach. Although our standardization efforts enhance performance within our specific context, this specificity might limit the applicability of our results to other scenarios or languages. As such, external validation would require extensive testing across diverse platforms and configurations to confirm the broader applicability of our findings.

\section{Conclusion}\label{sec:conclusion}

We present CG, a hybrid deep-learning model designed to detect poisoned source code samples.
We compared CG with the state-of-the-art approach ONION and found that CG significantly outperformed ONION. 
We also found that CG has a promising performance in uncovering new poison attacks.
CG can be integrated into development pipelines either as a preprocessing module to clean poisoned samples from training datasets or as a real-time module within an IDE to flag potential poisoning triggers.

Incorporating attention mechanisms to identify important features in code embeddings and understand how specific parts of the input code influence model decisions offers a promising direction for future research. Moreover, as large language models (LLMs) like GPT-4 gain prominence, exploring their potential to generate and detect poisoned samples using zero-shot and few-shot learning presents another valuable avenue for investigation.

\bibliographystyle{cas-model2-names}
\bibliography{paper-main}

\end{document}

%% file: macros.tex
\usepackage{ifthen}
\usepackage{xspace}
\usepackage[normalem]{ulem} 

\newboolean{showedits}
\setboolean{showedits}{true} 
\ifthenelse{\boolean{showedits}}
{
	\newcommand{\del}[1]{\textcolor{red}{\sout{#1}}} 
	\newcommand{\nbe}[3]{
		{\colorbox{#3}{\bfseries\sffamily\scriptsize\textcolor{white}{#1}}}
		{\textcolor{#3}{\sf\small$\blacktriangleright$\textit{#2}$\blacktriangleleft$}}}
}{
	\newcommand{\del}[1]{} 
	
	\newcommand{\nbe}[3]{}
}
\usepackage[most]{tcolorbox}
\ifthenelse{\boolean{showedits}}
{
  \newtcolorbox{inserted}{%
       title=Inserted text:,
       colframe=blue,colback=blue!5!white,
       breakable,
       leftrule=0mm, 
       bottomrule=0mm,
       rightrule=0mm,
       toprule=0mm,
       arc=0mm, outer arc=0mm,
       oversize
  }
  \newtcolorbox{deleted}{%
       title=Deleted text:,
       colframe=red,colback=red!5!white,
       breakable,
       leftrule=0mm, 
       bottomrule=0mm,
       rightrule=0mm,
       toprule=0mm,
       arc=0mm, outer arc=0mm,
       oversize
  }
  \newtcolorbox{refactored}{%
       title=Rewritten text:,
       colframe=blue,colback=red!5!white,
       breakable,
       leftrule=0mm, 
       bottomrule=0mm,
       rightrule=0mm,
       toprule=0mm,
       arc=0mm, outer arc=0mm,
       oversize
  }
}{

}
\newboolean{showcomments}
\setboolean{showcomments}{true}

\ifthenelse{\boolean{showcomments}}
{\newcommand{\nbc}[3]{
 {\colorbox{#3}{\bfseries\sffamily\scriptsize\textcolor{white}{#1}}}
 {\textcolor{#3}{\sf\small$\blacktriangleright$\textit{#2}$\blacktriangleleft$}}}
 }
{\newcommand{\nbc}[3]{}
 }

\newboolean{isblinded}
\setboolean{isblinded}{true}
\ifthenelse{\boolean{isblinded}}
{\newcommand\blind[1]{BLINDED\xspace}}
{\newcommand\blind[1]{#1\xspace}}

\usepackage{tcolorbox}
\newcommand{\boxit}[2][yellow!15]{%
    \begin{tcolorbox}[
        colback=#1,    
        colframe=black, 
        width=8.3cm,    
        arc=0mm,        
        boxrule=0.5pt,  
        fontupper=\small,
        left=2pt,       
        right=2pt,      
    ]
        \emph{#2}
    \end{tcolorbox}
}

%% file: paper-main.bbl
\begin{thebibliography}{38}
\expandafter\ifx\csname natexlab\endcsname\relax\def\natexlab#1{#1}\fi
\providecommand{\url}[1]{\texttt{#1}}
\providecommand{\href}[2]{#2}
\providecommand{\path}[1]{#1}
\providecommand{\DOIprefix}{doi:}
\providecommand{\ArXivprefix}{arXiv:}
\providecommand{\URLprefix}{URL: }
\providecommand{\Pubmedprefix}{pmid:}
\providecommand{\doi}[1]{\href{http://dx.doi.org/#1}{\path{#1}}}
\providecommand{\Pubmed}[1]{\href{pmid:#1}{\path{#1}}}
\providecommand{\bibinfo}[2]{#2}
\ifx\xfnm\relax \def\xfnm[#1]{\unskip,\space#1}\fi
\bibitem[{Aghakhani et~al.(2023)Aghakhani, Dai, Manoel, Fernandes, Kharkar, Kruegel, Vigna, Evans, Zorn and Sim}]{Aghakhani2023TrojanPuzzleCP}
\bibinfo{author}{Aghakhani, H.}, \bibinfo{author}{Dai, W.}, \bibinfo{author}{Manoel, A.}, \bibinfo{author}{Fernandes, X.}, \bibinfo{author}{Kharkar, A.}, \bibinfo{author}{Kruegel, C.}, \bibinfo{author}{Vigna, G.}, \bibinfo{author}{Evans, D.}, \bibinfo{author}{Zorn, B.G.}, \bibinfo{author}{Sim, R.}, \bibinfo{year}{2023}.
\newblock \bibinfo{title}{Trojanpuzzle: Covertly poisoning code-suggestion models}.
\newblock \bibinfo{journal}{2024 IEEE Symposium on Security and Privacy (SP)} , \bibinfo{pages}{1122--1140}\URLprefix \url{https://api.semanticscholar.org/CorpusID:255522506}.
\bibitem[{Alon et~al.(2018)Alon, Zilberstein, Levy and Yahav}]{alon2018code2vec}
\bibinfo{author}{Alon, U.}, \bibinfo{author}{Zilberstein, M.}, \bibinfo{author}{Levy, O.}, \bibinfo{author}{Yahav, E.}, \bibinfo{year}{2018}.
\newblock \bibinfo{title}{code2vec: Learning distributed representations of code}.
\newblock \href{http://arxiv.org/abs/1803.09473}{\tt arXiv:1803.09473}.
\bibitem[{Bojanowski et~al.(2017)Bojanowski, Grave, Joulin and Mikolov}]{bojanowski2017enriching}
\bibinfo{author}{Bojanowski, P.}, \bibinfo{author}{Grave, E.}, \bibinfo{author}{Joulin, A.}, \bibinfo{author}{Mikolov, T.}, \bibinfo{year}{2017}.
\newblock \bibinfo{title}{Enriching word vectors with subword information}.
\newblock \href{http://arxiv.org/abs/1607.04606}{\tt arXiv:1607.04606}.
\bibitem[{Bruni et~al.(2025)Bruni, Gabrielli, Ghafari and Kropp}]{Bruni2025}
\bibinfo{author}{Bruni, M.}, \bibinfo{author}{Gabrielli, F.}, \bibinfo{author}{Ghafari, M.}, \bibinfo{author}{Kropp, M.}, \bibinfo{year}{2025}.
\newblock \bibinfo{title}{Benchmarking prompt engineering techniques for secure code generation with gpt models}, in: \bibinfo{booktitle}{Proceedings of the 2025 IEEE/ACM Second International Conference on AI Foundation Models and Software Engineering}.
\bibitem[{Carlini and Wagner(2017)}]{carlini2017adversarial}
\bibinfo{author}{Carlini, N.}, \bibinfo{author}{Wagner, D.}, \bibinfo{year}{2017}.
\newblock \bibinfo{title}{Adversarial examples are not easily detected: Bypassing ten detection methods}, in: \bibinfo{booktitle}{Proceedings of the 10th ACM Workshop on Artificial Intelligence and Security}, \bibinfo{publisher}{Association for Computing Machinery}. p. \bibinfo{pages}{3–14}.
\newblock \URLprefix \url{https://doi.org/10.1145/3128572.3140444}, \DOIprefix\doi{10.1145/3128572.3140444}.
\bibitem[{Chen et~al.(2020)Chen, Salem, Chen, Backes, Ma, Shen, Wu and Zhang}]{Chen2020BadNLBA}
\bibinfo{author}{Chen, X.}, \bibinfo{author}{Salem, A.}, \bibinfo{author}{Chen, D.}, \bibinfo{author}{Backes, M.}, \bibinfo{author}{Ma, S.}, \bibinfo{author}{Shen, Q.}, \bibinfo{author}{Wu, Z.}, \bibinfo{author}{Zhang, Y.}, \bibinfo{year}{2020}.
\newblock \bibinfo{title}{Badnl: Backdoor attacks against nlp models with semantic-preserving improvements}.
\newblock \bibinfo{journal}{Proceedings of the 37th Annual Computer Security Applications Conference} \URLprefix \url{https://api.semanticscholar.org/CorpusID:238354397}.
\bibitem[{Chung et~al.(2014)Chung, Gulcehre, Cho and Bengio}]{chung2014empirical}
\bibinfo{author}{Chung, J.}, \bibinfo{author}{Gulcehre, C.}, \bibinfo{author}{Cho, K.}, \bibinfo{author}{Bengio, Y.}, \bibinfo{year}{2014}.
\newblock \bibinfo{title}{Empirical evaluation of gated recurrent neural networks on sequence modeling}.
\newblock \bibinfo{journal}{arXiv preprint arXiv:1412.3555} .
\bibitem[{Cotroneo et~al.(2024)Cotroneo, Improta, Liguori and Natella}]{cotroneo2024vulnerabilities}
\bibinfo{author}{Cotroneo, D.}, \bibinfo{author}{Improta, C.}, \bibinfo{author}{Liguori, P.}, \bibinfo{author}{Natella, R.}, \bibinfo{year}{2024}.
\newblock \bibinfo{title}{Vulnerabilities in ai code generators: Exploring targeted data poisoning attacks}, in: \bibinfo{booktitle}{Proceedings of the 32nd IEEE/ACM International Conference on Program Comprehension}, pp. \bibinfo{pages}{280--292}.
\bibitem[{Dou et~al.(2024)Dou, Wu, Jia, Zhou, Liu and Liu}]{dou2024cc2vec}
\bibinfo{author}{Dou, S.}, \bibinfo{author}{Wu, Y.}, \bibinfo{author}{Jia, H.}, \bibinfo{author}{Zhou, Y.}, \bibinfo{author}{Liu, Y.}, \bibinfo{author}{Liu, Y.}, \bibinfo{year}{2024}.
\newblock \bibinfo{title}{Cc2vec: Combining typed tokens with contrastive learning for effective code clone detection}.
\newblock \bibinfo{journal}{Proceedings of the ACM on Software Engineering} \bibinfo{volume}{1}, \bibinfo{pages}{1564--1584}.
\bibitem[{Feng et~al.(2020)Feng, Guo, Tang, Duan, Feng, Gong, Shou, Qin, Liu, Jiang and Zhou}]{feng2020codebert}
\bibinfo{author}{Feng, Z.}, \bibinfo{author}{Guo, D.}, \bibinfo{author}{Tang, D.}, \bibinfo{author}{Duan, N.}, \bibinfo{author}{Feng, X.}, \bibinfo{author}{Gong, M.}, \bibinfo{author}{Shou, L.}, \bibinfo{author}{Qin, B.}, \bibinfo{author}{Liu, T.}, \bibinfo{author}{Jiang, D.}, \bibinfo{author}{Zhou, M.}, \bibinfo{year}{2020}.
\newblock \bibinfo{title}{Codebert: A pre-trained model for programming and natural languages}.
\newblock \href{http://arxiv.org/abs/2002.08155}{\tt arXiv:2002.08155}.
\bibitem[{Firouzi et~al.(2024)Firouzi, Ghafari and Ebrahimi}]{Firouzi2024}
\bibinfo{author}{Firouzi, E.}, \bibinfo{author}{Ghafari, M.}, \bibinfo{author}{Ebrahimi, M.}, \bibinfo{year}{2024}.
\newblock \bibinfo{title}{Chatgpt’s potential in cryptography misuse detection: A comparative analysis with static analysis tools}, in: \bibinfo{booktitle}{Proceedings of the 18th ACM/IEEE International Symposium on Empirical Software Engineering and Measurement}, p. \bibinfo{pages}{582–588}.
\newblock \URLprefix \url{https://doi.org/10.1145/3674805.3695408}, \DOIprefix\doi{10.1145/3674805.3695408}.
\bibitem[{Group()}]{CoDesc}
\bibinfo{author}{Group, B.C.N.}, .
\newblock \bibinfo{title}{Codesc dataset}.
\newblock \URLprefix \url{https://github.com/csebuetnlp/CoDesc}. \bibinfo{note}{accessed on 2025-01-10}.
\bibitem[{Gu et~al.(2019)Gu, Dolan-Gavitt and Garg}]{gu2017badnets}
\bibinfo{author}{Gu, T.}, \bibinfo{author}{Dolan-Gavitt, B.}, \bibinfo{author}{Garg, S.}, \bibinfo{year}{2019}.
\newblock \bibinfo{title}{Badnets: Identifying vulnerabilities in the machine learning model supply chain}.
\newblock \URLprefix \url{https://arxiv.org/abs/1708.06733}, \href{http://arxiv.org/abs/1708.06733}{\tt arXiv:1708.06733}.
\bibitem[{Ji et~al.(2018)Ji, Zhang, Ji, Luo and Wang}]{Ji2018ModelReuseAO}
\bibinfo{author}{Ji, Y.}, \bibinfo{author}{Zhang, X.}, \bibinfo{author}{Ji, S.}, \bibinfo{author}{Luo, X.}, \bibinfo{author}{Wang, T.}, \bibinfo{year}{2018}.
\newblock \bibinfo{title}{Model-reuse attacks on deep learning systems}.
\newblock \bibinfo{journal}{Proceedings of the 2018 ACM SIGSAC Conference on Computer and Communications Security} \URLprefix \url{https://api.semanticscholar.org/CorpusID:53059573}.
\bibitem[{Jiang et~al.(2021)Jiang, Lutellier and Tan}]{jiang2021cure}
\bibinfo{author}{Jiang, N.}, \bibinfo{author}{Lutellier, T.}, \bibinfo{author}{Tan, L.}, \bibinfo{year}{2021}.
\newblock \bibinfo{title}{Cure: Code-aware neural machine translation for automatic program repair}, in: \bibinfo{booktitle}{2021 IEEE/ACM 43rd International Conference on Software Engineering (ICSE)}, \bibinfo{organization}{IEEE}. pp. \bibinfo{pages}{1161--1173}.
\bibitem[{Jin et~al.(2019)Jin, Jin, Zhou and Szolovits}]{jin2019bert}
\bibinfo{author}{Jin, D.}, \bibinfo{author}{Jin, Z.}, \bibinfo{author}{Zhou, J.T.}, \bibinfo{author}{Szolovits, P.}, \bibinfo{year}{2019}.
\newblock \bibinfo{title}{Is bert really robust? natural language attack on text classification and entailment}.
\newblock \bibinfo{journal}{arXiv preprint arXiv:1907.11932} .
\bibitem[{Jin et~al.(2022)Jin, Zhang, Shen, Chen, Fan, Lin and Liu}]{jin2022can}
\bibinfo{author}{Jin, K.}, \bibinfo{author}{Zhang, T.}, \bibinfo{author}{Shen, C.}, \bibinfo{author}{Chen, Y.}, \bibinfo{author}{Fan, M.}, \bibinfo{author}{Lin, C.}, \bibinfo{author}{Liu, T.}, \bibinfo{year}{2022}.
\newblock \bibinfo{title}{Can we mitigate backdoor attack using adversarial detection methods?}
\newblock \bibinfo{journal}{IEEE Transactions on Dependable and Secure Computing} \bibinfo{volume}{20}, \bibinfo{pages}{2867--2881}.
\bibitem[{Li et~al.(2024)Li, Li, Zhang, Li, Jin, Hu and Xia}]{Li2024}
\bibinfo{author}{Li, J.}, \bibinfo{author}{Li, Z.}, \bibinfo{author}{Zhang, H.}, \bibinfo{author}{Li, G.}, \bibinfo{author}{Jin, Z.}, \bibinfo{author}{Hu, X.}, \bibinfo{author}{Xia, X.}, \bibinfo{year}{2024}.
\newblock \bibinfo{title}{Poison attack and poison detection on deep source code processing models}.
\newblock \bibinfo{journal}{ACM Trans. Softw. Eng. Methodol.} \bibinfo{volume}{33}.
\newblock \URLprefix \url{https://doi.org/10.1145/3630008}, \DOIprefix\doi{10.1145/3630008}.
\bibitem[{Liu et~al.(2021)Liu, Qian, Wang, Zhuang, Qiu and Wang}]{liu2021combining}
\bibinfo{author}{Liu, Z.}, \bibinfo{author}{Qian, P.}, \bibinfo{author}{Wang, X.}, \bibinfo{author}{Zhuang, Y.}, \bibinfo{author}{Qiu, L.}, \bibinfo{author}{Wang, X.}, \bibinfo{year}{2021}.
\newblock \bibinfo{title}{Combining graph neural networks with expert knowledge for smart contract vulnerability detection}.
\newblock \bibinfo{journal}{IEEE Transactions on Knowledge and Data Engineering} \bibinfo{volume}{35}, \bibinfo{pages}{1296--1310}.
\bibitem[{Lu et~al.(2021)Lu, Guo, Ren, Huang, Svyatkovskiy, Blanco, Clement, Drain, Jiang, Tang, Li, Zhou, Shou, Zhou, Tufano, Gong, Zhou, Duan, Sundaresan, Deng, Fu and Liu}]{lu2021codexglue}
\bibinfo{author}{Lu, S.}, \bibinfo{author}{Guo, D.}, \bibinfo{author}{Ren, S.}, \bibinfo{author}{Huang, J.}, \bibinfo{author}{Svyatkovskiy, A.}, \bibinfo{author}{Blanco, A.}, \bibinfo{author}{Clement, C.}, \bibinfo{author}{Drain, D.}, \bibinfo{author}{Jiang, D.}, \bibinfo{author}{Tang, D.}, \bibinfo{author}{Li, G.}, \bibinfo{author}{Zhou, L.}, \bibinfo{author}{Shou, L.}, \bibinfo{author}{Zhou, L.}, \bibinfo{author}{Tufano, M.}, \bibinfo{author}{Gong, M.}, \bibinfo{author}{Zhou, M.}, \bibinfo{author}{Duan, N.}, \bibinfo{author}{Sundaresan, N.}, \bibinfo{author}{Deng, S.K.}, \bibinfo{author}{Fu, S.}, \bibinfo{author}{Liu, S.}, \bibinfo{year}{2021}.
\newblock \bibinfo{title}{Codexglue: A machine learning benchmark dataset for code understanding and generation}.
\newblock \href{http://arxiv.org/abs/2102.04664}{\tt arXiv:2102.04664}.
\bibitem[{Peng et~al.(2018)Peng, Li, He, Liu, Bao, Wang, Song and Yang}]{peng2018large}
\bibinfo{author}{Peng, H.}, \bibinfo{author}{Li, J.}, \bibinfo{author}{He, Y.}, \bibinfo{author}{Liu, Y.}, \bibinfo{author}{Bao, M.}, \bibinfo{author}{Wang, L.}, \bibinfo{author}{Song, Y.}, \bibinfo{author}{Yang, Q.}, \bibinfo{year}{2018}.
\newblock \bibinfo{title}{Large-scale hierarchical text classification with recursively regularized deep graph-cnn}, in: \bibinfo{booktitle}{Proceedings of the 2018 world wide web conference}, pp. \bibinfo{pages}{1063--1072}.
\bibitem[{QData(2023)}]{qdata_textattack}
\bibinfo{author}{QData}, \bibinfo{year}{2023}.
\newblock \bibinfo{title}{Textattack}.
\newblock \URLprefix \url{https://github.com/QData/TextAttack}. \bibinfo{note}{accessed on 2025-02-03}.
\bibitem[{Qi et~al.(2021)Qi, Chen, Li, Yao, Liu and Sun}]{qi-etal-2021-onion}
\bibinfo{author}{Qi, F.}, \bibinfo{author}{Chen, Y.}, \bibinfo{author}{Li, M.}, \bibinfo{author}{Yao, Y.}, \bibinfo{author}{Liu, Z.}, \bibinfo{author}{Sun, M.}, \bibinfo{year}{2021}.
\newblock \bibinfo{title}{{ONION}: A simple and effective defense against textual backdoor attacks}, in: \bibinfo{booktitle}{Proceedings of the 2021 Conference on Empirical Methods in Natural Language Processing}, \bibinfo{publisher}{Association for Computational Linguistics}, \bibinfo{address}{Online and Punta Cana, Dominican Republic}. pp. \bibinfo{pages}{9558--9566}.
\newblock \URLprefix \url{https://aclanthology.org/2021.emnlp-main.752}, \DOIprefix\doi{10.18653/v1/2021.emnlp-main.752}.
\bibitem[{Rawat and Wang(2017)}]{rawat2017deep}
\bibinfo{author}{Rawat, W.}, \bibinfo{author}{Wang, Z.}, \bibinfo{year}{2017}.
\newblock \bibinfo{title}{Deep convolutional neural networks for image classification: A comprehensive review}.
\newblock \bibinfo{journal}{Neural computation} \bibinfo{volume}{29}, \bibinfo{pages}{2352--2449}.
\bibitem[{Razmi and Xiong(2023)}]{razmi2023classification}
\bibinfo{author}{Razmi, F.}, \bibinfo{author}{Xiong, L.}, \bibinfo{year}{2023}.
\newblock \bibinfo{title}{Classification auto-encoder based detector against diverse data poisoning attacks}, in: \bibinfo{booktitle}{Data and Applications Security and Privacy XXXVII: 37th Annual IFIP WG 11.3 Conference, DBSec 2023, Sophia-Antipolis, France, July 19–21, 2023, Proceedings}, \bibinfo{publisher}{Springer-Verlag}, \bibinfo{address}{Berlin, Heidelberg}. p. \bibinfo{pages}{263–281}.
\newblock \URLprefix \url{https://doi.org/10.1007/978-3-031-37586-6_16}, \DOIprefix\doi{10.1007/978-3-031-37586-6_16}.
\bibitem[{Sahar et~al.(2024)Sahar, Younas, Khan and Sarwar}]{sahar2024dp}
\bibinfo{author}{Sahar, S.}, \bibinfo{author}{Younas, M.}, \bibinfo{author}{Khan, M.M.}, \bibinfo{author}{Sarwar, M.U.}, \bibinfo{year}{2024}.
\newblock \bibinfo{title}{Dp-ccl: A supervised contrastive learning approach using codebert model in software defect prediction}.
\newblock \bibinfo{journal}{IEEE Access} .
\bibitem[{Schuster et~al.(2020)Schuster, Song, Tromer and Shmatikov}]{Schuster2020YouAM}
\bibinfo{author}{Schuster, R.}, \bibinfo{author}{Song, C.}, \bibinfo{author}{Tromer, E.}, \bibinfo{author}{Shmatikov, V.}, \bibinfo{year}{2020}.
\newblock \bibinfo{title}{You autocomplete me: Poisoning vulnerabilities in neural code completion}, in: \bibinfo{booktitle}{USENIX Security Symposium}.
\newblock \URLprefix \url{https://api.semanticscholar.org/CorpusID:220363858}.
\bibitem[{Shafahi et~al.(2018)Shafahi, Huang, Najibi, Suciu, Studer, Dumitras and Goldstein}]{shafahi2018poison}
\bibinfo{author}{Shafahi, A.}, \bibinfo{author}{Huang, W.R.}, \bibinfo{author}{Najibi, M.}, \bibinfo{author}{Suciu, O.}, \bibinfo{author}{Studer, C.}, \bibinfo{author}{Dumitras, T.}, \bibinfo{author}{Goldstein, T.}, \bibinfo{year}{2018}.
\newblock \bibinfo{title}{Poison frogs! targeted clean-label poisoning attacks on neural networks}, in: \bibinfo{booktitle}{Proceedings of the 32nd International Conference on Neural Information Processing Systems}, \bibinfo{publisher}{Curran Associates Inc.}. p. \bibinfo{pages}{6106–6116}.
\bibitem[{Soarsmu(2023)}]{soarsmu_attack_pretrain_models_of_code}
\bibinfo{author}{Soarsmu}, \bibinfo{year}{2023}.
\newblock \bibinfo{title}{Attack pretrain models of code}.
\newblock \URLprefix \url{https://github.com/soarsmu/attack-pretrain-models-of-code}. \bibinfo{note}{accessed on 2025-02-03}.
\bibitem[{SRL(2023)}]{tech_srl_adversarial_examples}
\bibinfo{author}{SRL, T.}, \bibinfo{year}{2023}.
\newblock \bibinfo{title}{Adversarial examples}.
\newblock \URLprefix \url{https://github.com/tech-srl/adversarial-examples}. \bibinfo{note}{accessed on 2025-02-03}.
\bibitem[{Steinhardt et~al.(2017)Steinhardt, Koh and Liang}]{steinhardt2017certifieddefensesdatapoisoning}
\bibinfo{author}{Steinhardt, J.}, \bibinfo{author}{Koh, P.W.}, \bibinfo{author}{Liang, P.}, \bibinfo{year}{2017}.
\newblock \bibinfo{title}{Certified defenses for data poisoning attacks}, in: \bibinfo{booktitle}{Proceedings of the 31st International Conference on Neural Information Processing Systems}, \bibinfo{publisher}{Curran Associates Inc.}. p. \bibinfo{pages}{3520–3532}.
\bibitem[{Sun et~al.(2022)Sun, Du, Song, Ni and Li}]{sun2022coprotector}
\bibinfo{author}{Sun, Z.}, \bibinfo{author}{Du, X.}, \bibinfo{author}{Song, F.}, \bibinfo{author}{Ni, M.}, \bibinfo{author}{Li, L.}, \bibinfo{year}{2022}.
\newblock \bibinfo{title}{Coprotector: Protect open-source code against unauthorized training usage with data poisoning}, in: \bibinfo{booktitle}{Proceedings of the ACM Web Conference 2022}, pp. \bibinfo{pages}{652--660}.
\bibitem[{Wan et~al.(2022)Wan, Zhang, Zhang, Sui, Xu, Yao, Jin and Sun}]{Wan2022YouSW}
\bibinfo{author}{Wan, Y.}, \bibinfo{author}{Zhang, S.}, \bibinfo{author}{Zhang, H.}, \bibinfo{author}{Sui, Y.}, \bibinfo{author}{Xu, G.}, \bibinfo{author}{Yao, D.}, \bibinfo{author}{Jin, H.}, \bibinfo{author}{Sun, L.}, \bibinfo{year}{2022}.
\newblock \bibinfo{title}{You see what i want you to see: poisoning vulnerabilities in neural code search}.
\newblock \bibinfo{journal}{Proceedings of the 30th ACM Joint European Software Engineering Conference and Symposium on the Foundations of Software Engineering} \URLprefix \url{https://api.semanticscholar.org/CorpusID:253421850}.
\bibitem[{Wang et~al.(2020)Wang, Li, Ma, Xia and Jin}]{wang2020detecting}
\bibinfo{author}{Wang, W.}, \bibinfo{author}{Li, G.}, \bibinfo{author}{Ma, B.}, \bibinfo{author}{Xia, X.}, \bibinfo{author}{Jin, Z.}, \bibinfo{year}{2020}.
\newblock \bibinfo{title}{Detecting code clones with graph neural network and flow-augmented abstract syntax tree}, in: \bibinfo{booktitle}{2020 IEEE 27th International Conference on Software Analysis, Evolution and Reengineering (SANER)}, \bibinfo{organization}{IEEE}. pp. \bibinfo{pages}{261--271}.
\bibitem[{Yang et~al.(2022)Yang, Shi, He and Lo}]{Yang_2022}
\bibinfo{author}{Yang, Z.}, \bibinfo{author}{Shi, J.}, \bibinfo{author}{He, J.}, \bibinfo{author}{Lo, D.}, \bibinfo{year}{2022}.
\newblock \bibinfo{title}{Natural attack for pre-trained models of code}, in: \bibinfo{booktitle}{Proceedings of the 44th International Conference on Software Engineering}, \bibinfo{publisher}{{ACM}}.
\newblock \URLprefix \url{https://doi.org/10.1145%2F3510003.3510146}, \DOIprefix\doi{10.1145/3510003.3510146}.
\bibitem[{Yang et~al.(2024)Yang, Xu, Zhang, Kang, Shi, He and Lo}]{Yang2024}
\bibinfo{author}{Yang, Z.}, \bibinfo{author}{Xu, B.}, \bibinfo{author}{Zhang, J.M.}, \bibinfo{author}{Kang, H.J.}, \bibinfo{author}{Shi, J.}, \bibinfo{author}{He, J.}, \bibinfo{author}{Lo, D.}, \bibinfo{year}{2024}.
\newblock \bibinfo{title}{Stealthy backdoor attack for code models}.
\newblock \bibinfo{journal}{IEEE Trans. Softw. Eng.} \bibinfo{volume}{50}, \bibinfo{pages}{721–741}.
\newblock \URLprefix \url{https://doi.org/10.1109/TSE.2024.3361661}, \DOIprefix\doi{10.1109/TSE.2024.3361661}.
\bibitem[{Yefet et~al.(2020)Yefet, Alon and Yahav}]{yefet2020adversarial}
\bibinfo{author}{Yefet, N.}, \bibinfo{author}{Alon, U.}, \bibinfo{author}{Yahav, E.}, \bibinfo{year}{2020}.
\newblock \bibinfo{title}{Adversarial examples for models of code}.
\newblock \bibinfo{journal}{Proc. ACM Program. Lang.} \bibinfo{volume}{4}.
\newblock \URLprefix \url{https://doi.org/10.1145/3428230}, \DOIprefix\doi{10.1145/3428230}.
\bibitem[{Zhang et~al.(2020)Zhang, Li, Li, Ma, Liu and Jin}]{zhang2020generating}
\bibinfo{author}{Zhang, H.}, \bibinfo{author}{Li, Z.}, \bibinfo{author}{Li, G.}, \bibinfo{author}{Ma, L.}, \bibinfo{author}{Liu, Y.}, \bibinfo{author}{Jin, Z.}, \bibinfo{year}{2020}.
\newblock \bibinfo{title}{Generating adversarial examples for holding robustness of source code processing models}, in: \bibinfo{booktitle}{Proceedings of the AAAI Conference on Artificial Intelligence}, pp. \bibinfo{pages}{1169--1176}.

\end{thebibliography}
